\documentclass[11pt]{article}
\usepackage{hyperref}
\hypersetup{colorlinks=true}
\hypersetup{linkcolor=black}
\hypersetup{citecolor=black}
\hypersetup{urlcolor=black}
\usepackage{slashed}

\usepackage[left=2.5cm,right=2.5cm,top=2.5cm,bottom=3cm]{geometry}
\usepackage{tikz}
\usetikzlibrary{arrows,decorations.markings,cd}
\tikzset{
->-/.style={decoration={markings, mark=at position .5 with {\arrow[scale=1.5]{stealth}}}, postaction={decorate}}
}
\usepackage{epsfig}
\usepackage{amssymb}
\usepackage{amsfonts}
\usepackage{amsbsy}
\usepackage{amsmath}
\usepackage{dsfont}
\usepackage{bbm}
\usepackage{upgreek}
\usepackage{amscd}
\usepackage[nosort]{cite}
\usepackage{graphicx}
\usepackage{mathrsfs}
\usepackage{amsmath,amsthm}
\usepackage{slashed}	
\usepackage{dsfont}
\usepackage{tikz}
\usepackage[utf8]{inputenc}
\usepackage{tikz-feynman}\tikzfeynmanset{compat=1.0.0}
\numberwithin{equation}{section}

\usetikzlibrary{positioning}
\usetikzlibrary{shapes.geometric, arrows}
\tikzstyle{rect} = [rectangle,rounded corners,minimum width=3cm, minimum height=1cm, text centered, draw=black]
\tikzstyle{arrow} = [thick,->,>=stealth]
\usepackage{stackrel}
\tikzstyle{oct} = [regular polygon,regular polygon sides=8, draw,
    text width=2em, text centered]
\tikzstyle{line} = [draw, -latex']

\tikzstyle{hex} = [regular polygon,regular polygon sides=6, draw,
    text width=2em, text centered]
\tikzstyle{hexs}= [regular polygon,regular polygon sides=5, draw,
    text width=1.7em, text centered]

\makeatletter
\DeclareFontFamily{OMX}{MnSymbolE}{}
\DeclareSymbolFont{MnLargeSymbols}{OMX}{MnSymbolE}{m}{n}
\SetSymbolFont{MnLargeSymbols}{bold}{OMX}{MnSymbolE}{b}{n}
\DeclareFontShape{OMX}{MnSymbolE}{m}{n}{
    <-6>  MnSymbolE5
   <6-7>  MnSymbolE6
   <7-8>  MnSymbolE7
   <8-9>  MnSymbolE8
   <9-10> MnSymbolE9
  <10-12> MnSymbolE10
  <12->   MnSymbolE12
}{}
\DeclareFontShape{OMX}{MnSymbolE}{b}{n}{
    <-6>  MnSymbolE-Bold5
   <6-7>  MnSymbolE-Bold6
   <7-8>  MnSymbolE-Bold7
   <8-9>  MnSymbolE-Bold8
   <9-10> MnSymbolE-Bold9
  <10-12> MnSymbolE-Bold10
  <12->   MnSymbolE-Bold12
}{}

\let\llangle\@undefined
\let\rrangle\@undefined
\DeclareMathDelimiter{\llangle}{\mathopen}%
                     {MnLargeSymbols}{'164}{MnLargeSymbols}{'164}
\DeclareMathDelimiter{\rrangle}{\mathclose}%
                     {MnLargeSymbols}{'171}{MnLargeSymbols}{'171}
\makeatother

\def\be{ \begin{equation} }
\def\ee{ \end{equation}}

\def\exp{{\rm exp}}

\def\mod{{\rm mod}}

\def\half{\frac{1}{2}}

\def\one{{\hbox{ 1\kern-.8mm l}}}

\def\CA{{\cal A}}

\def\CN {{\cal N}}

\def\CO {{\cal O}}

\def\CO {{\cal O}}

\def\IR{{\mathbb{R}}}

\def\IZ{{\mathbb{Z}}}

\def\fe{\mathfrak{e}}

\def\rmk#1{\bigskip\noindent{\bf Remark} }
\def\cnj#1{\bigskip\noindent{\bf Conjecture:} }

\DeclareMathAlphabet{\mathpzc}{OT1}{pzc}{m}{it}

\def\Tr{ \, \textrm{Tr} \, }

\def\t{\tilde}

\DeclareFontShape{OT1}{cmr}{mx}{n}%
    {<->cmr10}{}
\newcommand{\mytitlefont}{\fontseries{mx}\selectfont}
\DeclareMathAlphabet{\titlemath}{OT1}{cmr}{mx}{n}

\def\ie{\begin{equation}\begin{aligned}}
\def\fe{\end{aligned}\end{equation}}

\begin{document}

\begin{titlepage}

\begin{center}

~\\[2cm]

{\fontsize{24pt}{0pt} \mytitlefont  Line Defect Quantum Numbers \& Anomalies}

~\\[0.1cm]

T. Daniel Brennan,$^{1,2}$ Clay C\'{o}rdova,$^1$ and Thomas T.~Dumitrescu\,$^3$ 

~\\[0.1cm]

$^1$\,{\it Kadanoff Center for Theoretical Physics \& Enrico Fermi Institute, University of Chicago}\\[4pt]

$^2$\,{\it School of Mathematics, University of Birmingham
Watson Building,}\\[-2pt]
{\it Edgbaston, Birmingham B15 2TT, UK}\\[4pt]

$^3$\,{\it Mani L.\,Bhaumik Institute for Theoretical Physics, Department of Physics and Astronomy,}\\[-2pt]
       {\it University of California, Los Angeles, CA 90095, USA}\\[4pt]

~\\[15pt]

\end{center}

\noindent We explore the connection between the global symmetry quantum numbers of line defects and 't~Hooft anomalies. Relative to local (point) operators, line defects may transform projectively under both internal and spacetime symmetries. This phenomenon is known as symmetry fractionalization, and in general it signals the presence of certain discrete 't Hooft anomalies.  We describe this in detail in the context of free Maxwell theory in four dimensions. This understanding allows us to deduce the 't Hooft anomalies of non-Abelian gauge theories with renormalization group flows into Maxwell theory by analyzing the fractional quantum numbers of dynamical magnetic monopoles. We illustrate this method in $SU(2)$ gauge theories with matter fermions in diverse representations of the gauge group. For adjoint matter, we uncover a mixed anomaly involving the 0-form and 1-form symmetries, extending previous results. For~$SU(2)$ QCD with fundamental fermions, the 't~Hooft anomaly for the 0-form symmetries that is encoded by the fractionalization patterns of lines in the Maxwell phase is a consequence of the familiar perturbative (triangle) anomaly. 

\vfill

\begin{flushleft}
June 2022
\end{flushleft}

\end{titlepage}

\setcounter{tocdepth}{3}
	
\tableofcontents

\section{Introduction}
\label{intro}

In this paper we explore the connection between two remarkable phenomena in quantum field theory:    
\begin{itemize}
\item \emph{Symmetry fractionalization}:  the discrepancy between the global symmetry quantum numbers of solitons or defects and those of local (point) operators.  For instance, as first discovered by Jackiw and Rebbi~\cite{Jackiw:1975fn}, magnetic monopoles in gauge theory often have fractional flavor quantum numbers.  Another famous example is the fractional quantum hall effect, where quasiparticles, modeled by line defects in a topological field theory, have a fractional charge.

\item \emph{'t Hooft anomalies}: obstructions to gauging the global symmetry~\cite{tHooft:1979rat}.  In modern terms, such anomalies are characterized by anomaly inflow \cite{Callan:1984sa} from a classical local action (also called an invertible field theory \cite{Freed:2004yc, Freed:2016rqq} or symmetry protected topological phase) in one higher spacetime dimension.  Anomalies are a fundamental tool in studying the dynamics of field theory due to their invariance under continuous symmetry-preserving deformations, including in particular renormalization group flows.  
\end{itemize}
We explain why these two concepts are closely related and show that symmetry fractionalization of line defects often indicates the presence of subtle discrete 't Hooft anomalies.  Thus, carefully analyzing the quantum numbers of line defects is a systematic physical way to deduce certain global anomalies in QFT.  We illustrate this approach in a simple class of non-Abelian gauge theory examples.  We comment on some of the many possible generalizations at the end of this introduction.

Both symmetry fractionalization and 't Hooft anomalies have been intensely studied -- in high-energy physics, string theory, condensed matter physics, and mathematics -- and many applications and generalizations have been found. 
In three spacetime dimensions \cite{Barkeshli:2014cna, Barkeshli:2017rzd, Barkeshli:2019vtb, Hsin:2019gvb} pioneered a systematic approach to symmetry fractionalization in topological field theory; more recently \cite{Cordova:2017kue, Cheng:2022nji, Bhardwaj:2022dyt} have emphasized the relationship of between fractionalization and anomalies.  In four spacetime dimensions (the context of the present paper) \cite{Hsin:2019fhf, Ang:2019txy} studied symmetry fractionalization, while \cite{Thorngren:2014pza, Kravec:2014aza, Wang:2018qoy} described a concrete example (discussed in detail below) of how symmetry fractionalization for magnetic monopoles can lead to an anomaly.  Closely related examples have been explored in \cite{Cordova:2018acb, Wang:2019obe, Lee:2021crt}.\footnote{~Recently, \cite{Sato:2022vii} studied the properties of magnetic monopoles as an indicator of anomalies that constitute an inconsistency in the theory (e.g.~a gauge anomaly).  Our focus here is the connection between monopoles and 't Hooft anomalies for global symmetries, which are not inconsistencies, but important observables. Nevertheless, the criterion advocated in \cite{Sato:2022vii} will naturally make an appearance in our discussion of monopoles.}

In the context of four-dimensional gauge theory, it is natural to discuss the -- generally fractionalized --  quantum numbers of Wilson and 't Hooft line defects under a global 0-form symmetry~$G^{(0)}$.\footnote{~Here we are using the terminology of~\cite{Gaiotto:2014kfa}: a $p$-form global symmetry $G^{(p)}$ has charged objects that are extended operators of dimension $p$. See~\cite{Cordova:2022ruw} for a recent summary with references.  In general we use~$X^{(p)}$ to emphasize that~$X$ is a $p$-form.} How exactly these quantum numbers are to be understood is discussed below. In general the line defects represent infinitely heavy probe particles, but a subset of them may also correspond to dynamical charged particles in the spectrum of the theory. The main idea that we would like to explore is the relation between the~$G^{(0)}$ quantum numbers of Wilson and 't Hooft lines and 't Hooft anomalies involving the~$G^{(0)}$ symmetry.  For simplicity, throughout our discussion we will assume that~$G^{(0)}$ is a connected (but typically not simply-connected) group. 

The hero of our story will be free Maxwell theory in four dimensions (without charged matter). It is a~$U(1)$ gauge theory formulated using a dynamical~$U(1)$ connection~$a^{(1)}$ with field strength~$f^{(2)} = da^{(1)}.$ Maxwell theory is bosonic, i.e.~its Euclidean spacetime symmetry is~$SO(4)$, and hence it can be studied on arbitrary oriented Riemannian four-manifolds~$M_4$.\footnote{~In this paper we will not assume time-reversal symmetry, which is the reason we restrict to oriented~$M_4$, so that~$w_1(M_4) = 0$. Here~$w_n(M_4) \in H^n(M_4, \IZ_2)$ denotes the $n$-th Stiefel-Whitney classes of the tangent bundle of~$M_4$.} In particular, $M_4$ need not admit a spin structure, in which case~$w_2(M_4) \neq 0$. For simplicity, we assume that the Maxwell theory has vanishing~$\theta$-angle.

Maxwell theory does not possess intrinsic continuous 0-form symmetries that act faithfully on its local operators. However, it does have intrinsic electric ($e$) and magnetic ($m$) 1-form symmetries~\cite{Gaiotto:2014kfa},
\begin{equation}\label{max1fm}
U(1)_e^{(1)} \times U(1)_m^{(1)}~.
\end{equation}
The associated conserved 1-form charges are the quantized electric and magnetic charges $(E, M) \in \IZ \times \IZ$ of Wilson-'t Hooft lines. An important role will be played by the 2-form background gauge fields~$B_{e,m}^{(2)}$ for the 1-form symmetry~\eqref{max1fm}, which couple to the electric and magnetic flux densities.  Crucially, in the presence of both background fields, Maxwell theory has an 't Hooft anomaly which can be cancelled via inflow from a five-manifold~$M_5$ with boundary~$ \partial M_5 = M_4$ and the following action for the background fields~\cite{Gaiotto:2014kfa},
\begin{equation}\label{bebmanomaly}
\mathcal{A}= {i \over 2\pi} \int_{M_5} B_m^{(2)} \wedge d B_e^{(2)}~.
\end{equation}

In most of our applications we focus on discrete subgroups of the 1-form symmetries, the simplest case being
\begin{equation}
\IZ_{2,e}^{(1)} \times \IZ_{2,m}^{(1)} \subset U(1)_e^{(1)} \times U(1)_m^{(1)}~. 
\end{equation}
The appropriate background gauge fields are~$b^{(2)}_{e,m} \in H^2(M_4, \IZ_2)$, which couple to Maxwell theory via flat~$B$-fields of the form~$B_{e,m}^{(2)} = \pi b_{e,m}^{(2)}$. Specializing the 't Hooft anomaly~\eqref{bebmanomaly} to these background fields reduces it to a~$\IZ_2$-valued anomaly,
\begin{equation}\label{z2anom}
\mathcal{A} = i \pi \int_{M_5} b_m^{(2)} \cup \beta_2(b_e^{(2)})~.
\end{equation}
Here~$\beta_2(b_e^{(2)}) \in H^3(M_5, \IZ_2)$ is a~$\IZ_2$ analogue of the integral class~${1 \over 2\pi} d B_e^{(2)}$, with the Bockstein map~$\beta_2$ playing the role of the exterior derivative, while the wedge product is replaced by a cup product for discrete cohomology. (See the discussion around \eqref{bockdef} for additional details.)

We now seek to enrich Maxwell theory by a~$G^{(0)}$ symmetry, paying careful attention to the global form of the symmetry group. Since we have assumed that $G^{(0)}$ is connected it leaves the charges $(E,M)$ invariant.  To consistently define the fractionalized~$G^{(0)}$ quantum numbers of the lines we now make the additional simplifying assumption that $G^{(0)}$ is not part of a non-trivial 2-group global symmetry (see for instance~\cite{Tachikawa:2017gyf,Cordova:2018cvg,Benini:2018reh}) with the electric or magnetic 1-form symmetries.\footnote{~If~$G^{(0)}$ participates in a two-group together with a 1-form global symmetry~$G^{(1)}$ and non-trivial Postnikov class in~$H^3(G^{(0)}, G^{(1)})$ then lines charged under~$G^{(1)}$ cannot be assigned well-defined fractionalized~$G^{(0)}$ quantum numbers. For this reason, a 2-group with non-vanishing Postnikov class is often referred to as an obstruction to symmetry fractionalization in the condensed matter literature, see for instance \cite{Barkeshli:2014cna}, as well as \cite{Bhardwaj:2021wif, Lee:2021crt, Apruzzi:2021mlh, Bhardwaj:2022dyt} for recent field theory discussions.} A prototypical example is~$G^{(0)} = SO(3)$. Given our assumptions, there are two possibilities for the~$G^{(0)}$ quantum numbers of lines in Maxwell theory:

\begin{itemize}
\item[(i)] A line~$L(C)$, extended along a curve~$C$, transforms in a genuine representation~$\mathbf{R}$ of~$G^{(0)}$, i.e.~an integer-spin representation of~$SO(3)$. This is captured by placing a Wilson line in representation~$\mathbf{R}$ for the~$G^{(0)}$ background gauge field~$A^{(1)}$ along~$C$, which multiplies~$L(C)$ by a c-number that is local along the curve $C$ supporting the line,
\begin{equation}
L(C) \rightarrow L(C) \, \text{Tr}_{\mathbf{R}} \left( \exp\left(i \int_C A^{(1)}\right) \right)~.
\end{equation}
Such a background Wilson line should be thought of as a local counterterm (analogous to wave-function renormalization for local operators), hence it is scheme dependent and therefore UV sensitive. 

\item[(ii)] The line~$L(C)$ can transform projectively under~$G^{(0)} = SO(3)$, i.e.~it can transform in a half-integer representation of the universal covering group~$SU(2)$. In this case the symmetry is fractionalized. Since the choice of projective representation is only physical modulo genuine non-projective representations, it is completely captured by the charge of the line under the~$\IZ_2$ center of~$SU(2)$ which we mod out by to get to~$SO(3)$. 

In order to indicate that~$L(C)$ transforms non-trivially under this~$\IZ_2$, we describe the projective representation as an~$SO(3)$ 't Hooft anomaly for the one-dimensional quantum mechanics living on the line. This can be done by attaching to~$L(C)$ an anomaly inflow action that lives on a two-dimensional sheet~$\Sigma_{2}$ with boundary~$C$,
\begin{equation}
L(C) \rightarrow L(C)\, \exp\left(i \pi \int_{\Sigma_{2}} w_2(SO(3))\right)~, \qquad \partial\Sigma_{2} =C~.
\end{equation}
Here~$w_2(SO(3)) \in H^2(M_4, \IZ_2)$ is the second Stiefel-Whitney class of the~$SO(3)$ bundle, which obstructs its lift to an~$SU(2)$ bundle. 

\end{itemize}

In Maxwell theory, the background fields~$b_{e,m}^{(2)} \in H^2(M_4, \IZ_2)$ allow us to fractionalize the Wilson or the 't Hooft line under the~$SO(3)$ symmetry, because they automatically attach sheets of the form~$\exp\left(i \pi \int_{\Sigma_{2}} b_{e,m}^{(2)}\right)$ to the respective lines. We thus arrive at the following prescription:
\begin{itemize}
\item[(1.)] If the fundamental~$(E,M) = (1,0)$ Wilson line transforms projectively under~$SO(3)$, we set
\begin{equation}
b_e^{(2)} = w_2(SO(3))~.
\end{equation}
\item[(2.)] If the fundamental~$(E,M) = (0,1)$ 't Hooft line transforms projectively under~$SO(3)$, we activate
\begin{equation}
b_m^{(2)} = w_2(SO(3))~.
\end{equation}

\end{itemize}
Comparing with~\eqref{z2anom}, we immediately conclude that if both the Wilson and the 't Hooft lines transform projectively under~$SO(3)$, there is an 't Hooft anomaly of the form
\begin{equation}\label{so3bso3anom}
\mathcal{A} = {i \pi} \int_{M_5} w_2(SO(3)) \cup \beta_2\left(w_2(SO(3))\right) = i \pi \int_{M_5} w_2(SO(3)) \cup w_3(SO(3))~.
\end{equation}
Thus certain 't Hooft anomalies can be detected by analyzing the fractionalized charges of line defects -- in this case the Wilson and 't Hooft lines of free Maxwell theory.

In section~\ref{maxlines} we review Maxwell theory in more detail, and we explain how to account for the fractionalization of any connected 0-form symmetry~$G^{(0)}$. This includes the fractionalization of spin in bosonic theories (for which~$M_4$ need not be a spin manifold), which is described by a two-dimensional anomaly inflow action involving~$w_2(M_4),$ the second Stiefel-Whitney class of the tangent bundle.  In this way we realize a variety of possible discrete 't Hooft anomalies which may be physically detected by symmetry fractionalization.

Discrete anomalies of the general form \eqref{so3bso3anom}, involving discrete 2-form background fields arising from 0-form or 1-form symmetries have recently appeared in a variety of contexts including the study of strongly-coupled gauge theories and grand unification \cite{Wan:2018bns, Wan:2018djl, Wan:2018zql, Wan:2019oyr, Wan:2020ynf, Wan:2019oax,  Anber:2020gig, Wang:2021hob,Anber:2021iip}. The connection of such anomalies with monopole quantum numbers in non-Abelian gauge theories was utilized previously in \cite{Thorngren:2014pza, Cordova:2018acb, Wang:2018qoy, Lee:2021crt}.  A notable example is the anomaly of all-fermion electrodynamics \cite{Thorngren:2014pza, Kravec:2014aza}, i.e.~Maxwell theory where the fundamental~$(1,0)$ Wilson and~$(0,1)$ 't Hooft lines, as well as the~$(1,1)$ dyon obtained by fusing them, are fractionalized with respect to spacetime rotations, because they are fermionic lines in a bosonic QFT. This gives a simple four-dimensional field theory with a discrete gravitational anomaly (see section \ref{sec:allfed}, as well as~\cite{Chen:2021xks, Fidkowski:2021unr, Wan:2021idn} for further discussion.) 

The invariance of 't Hooft anomalies under deformations implies that we can apply the preceding discussion to learn something about the anomalies of any theory that can be connected to pure Maxwell theory by an RG flow that preserves some~$G^{(0)}$ symmetry. One physical context where such RG flows arise naturally is in $\mathcal{N}=2$ supersymmetric QFTs deformed onto their Coulomb branch, which is described by an ${\cal N} = 2$ supersymmetric version of Maxwell theory.  For instance, an example of a theory that possesses the anomaly~\eqref{so3bso3anom} is the original Argyres-Douglas SCFT \cite{Argyres:1995jj, Argyres:1995xn}, where the~$G^{(0)} = SO(3)$ symmetry is covered by the~$SU(2)_R$ symmetry under which the~${\cal N} = 2$ supercharges transform as a doublet.\footnote{~As explained in section~\ref{su2adjoints}, the Argyres-Douglas SCFT is fermionic, with spacetime symmetry~$\text{Spin}(4)$, but~$(-1)^F$ is identified with the central element~$(-1) \in SU(2)_R$. Thus the full connected symmetry is~$G^{(0)} = (SU(2)_R \times \text{Spin}(4))/\IZ_2$ and there is a single class~$w_2(G^{(0)}) \in H^2(M_4, \IZ_2)$ that obstructs lifts of~$G^{(0)}$-bundles to bundles of the covering group~$SU(2)_R \times \text{Spin}(4)$. This is the class that appears in the anomaly~\eqref{so3bso3anom}.} 

In section~\ref{rgflows} we outline a systematic approach to studying certain 't Hooft anomalies of non-Abelian gauge theories with~$G^{(0)}$ symmetry by adding adjoint Higgs fields and Yukawa couplings that allow us to trigger~$G^{(0)}$-preserving RG flows to pure Abelian gauge theory phases.\footnote{~Note that~$G^{(0)}$ is typically a subgroup of the theory without Yukawa couplings. Moreover, the ability to flow to pure Abelian gauge theory via the addition of Yukawa couplings restricts the possible matter content beyond what is required by gauge anomaly cancellation alone. } We then determine the fractionalized~$G^{(0)}$ quantum numbers of Abelian Wilson and 't Hooft lines -- and hence the~$G^{(0)}$ 't Hooft anomaly -- by examining the massive electrically and magnetically charged particles in the full theory. This is straightforward for Wilson lines, but as discovered in~\cite{Jackiw:1975fn} magnetic monopoles (and hence the Abelian 't Hooft lines of the IR theory) can transform in interesting projective representations of~$G^{(0)}$ due to the presence of fermion zero modes bound to the monopoles. 

An important conceptual point about this paradigm is that the 1-form global symmetry \eqref{max1fm} of the IR Maxwell theory is typically emergent along the RG flow, because the UV parent theory often does not enjoy such a symmetry. (And even if the UV theory does possess some 1-form symmetry, it is typically much smaller than \eqref{max1fm}.) Nevertheless, the emergent 1-form symmetry is crucial for correctly implementing the notion of symmetry fractionalization of the~$G^{(0)}$ symmetry in the Maxwell phase, and to correctly match the 't Hooft anomalies of this symmetry. 

In order to illustrate the method sketched above, we explore in detail the case of~$SU(2)$ gauge theories with matter fermions in diverse representations of the gauge group. The details of this analysis can be found in section~\ref{rgflows}. Here we summarize some of the highlights:
\begin{itemize}
\item We study~$SU(2)$ gauge theory with a single Weyl fermion in the~$D$-dimensional representation~$\bf D$ of the gauge group. The ability to deform to pure Maxwell theory using an adjoint Higgs field and Yukawa couplings requires~$D$ to be even, and the absence of gauge anomalies of Witten type \cite{Witten:1982fp} requires~$D \in 4 \IZ$. As we will review, these theories are bosonic, because~$(-1)^F$ fermion parity is gauged, and hence they can be studied on orientable manifolds with~$w_2(M_4) \neq 0$. This means the connected symmetry is~$G^{(0)} = SO(4)$.  We will show that this theory has a gravitational 't Hooft anomaly of the form
\begin{equation}
\mathcal{A} = {i \pi D \over 4} \int_{M_5} w_2(M_5) \cup w_3(M_5)~,
\end{equation}
which is non-trivial when~$D \equiv 4 ~ (\text{mod } 8)$.  This reproduces a result of \cite{Wang:2018qoy} and shows in particular that for $D \equiv 4 ~ (\text{mod } 8)$, this model flows to all-fermion electrodynamics  \cite{Thorngren:2014pza, Kravec:2014aza}.  

\item We study~$SU(2)$ gauge theory with~$N_f = 2 n_f$ Weyl fermions (or~$n_f$ Dirac fermions) in the fundamental representation of~$SU(2)$. In the presence of Yukawa couplings, the connected 0-form symmetry of this theory is
\begin{equation}
G^{(0)} = {SO(2n_f) \over \IZ_2} \times SO(4)~.
\end{equation} 
By studying the fractionalized quantum numbers of Wilson and 't Hooft lines in the IR Maxwell theory, we show that this theory has an 't Hooft anomaly. This anomaly is easiest to state for the special case where we restrict the flavor backgrounds to bundles of $SO(2n_{f})$, in which case it takes the form
\begin{equation}\label{fundanom}
\mathcal{A} = i \pi \int_{M_5} w_{2}\left(M_{5}\right) \cup \beta_2\left( w_{2}(SO(2n_f))\right) ~,
\end{equation}
for all~$n_f$. Here~$w_{2}(SO(2n_f)) \in H^2(M_5, \IZ_2)$ is the obstruction to lifting the flavor bundle to $\text{Spin}(2n_{f})$.  We also derive the anomaly for the richer class of $SO(2n_f) / \IZ_2$ flavor bundles where \eqref{fundanom} is further refined and depends on~$n_f \text{ mod } 2$.  An interesting feature of the anomaly in~\eqref{fundanom} is that it descends from the perturbative chiral (i.e.\ triangle) anomaly for the~$SU(2n_f)/\IZ_2$ flavor symmetry, which is broken to $SO(2n_f)/\IZ_2$ by the Yukawa couplings.

\item We study~$SU(2)$ gauge theory with~$N_f = 2n_f$ Weyl fermions in the adjoint representation of~$SU(2)$. In the presence of Yukawa couplings, the connected 0-form symmetry is
\begin{equation}
G^{(0)} = {USp(2n_f) \times \text{Spin}(4) \over \IZ_2}~.
\end{equation}
Since the adjoint matter fields do not transform under the center of the~$SU(2)$ gauge group, the theory also has a~$\IZ_2^{(1)}$ electric 1-form symmetry that protects the fundamental Wilson line. We show that this theory has an 't Hooft anomaly of the form
\begin{equation}
\mathcal{A} = i \pi n_f \int_{M_5} B_{\text{UV}}^{(2)} \cup \beta_2 \left(w_2(G^{(0)})\right)~. 
\end{equation}
Here~$w_2(G^{(0)}) \in H^2(M_5, \IZ_2)$ is the obstruction to lifting a~$G^{(0)}$ bundle to a~$USp(2n_f) \times \text{Spin}(4)$ bundle, while $B_{\text{UV}}^{(2)}\in H^2(M_5, \IZ_2)$ is the $\mathbb{Z}_{2}^{(1)}$ background field (extended to~$M_5$). Specializing to~$n_f = 1$ reproduces the anomaly first discussed in \cite{Cordova:2018acb}, originally deduced from computations in (topologically twisted) $\mathcal{N}=2$ supersymmetric pure~$SU(2)$ gauge theory~\cite{Witten:1994cg,Witten:1995gf, Moore:1997pc} (see also \cite{Wang:2018qoy}).  Note that the case $n_{f}=2$ (i.e.~$N_f = 4$ adjoint fermions) is connected to the physics of $\mathcal{N}=4$ supersymmetric $SU(2)$ gauge theory on its Coulomb branch.

\end{itemize}

The discussion in the present paper can be generalized in many directions, some of which are being actively explored in the literature.  This includes disconnected symmetry groups (e.g.~time-reversal symmetry \cite{Wang:2015fmi, Barkeshli:2017rzd, Zou:2017ppq, Hsin:2019fhf, Wang:2019obe}), as well as higher-rank gauge theories, or non-Lagrangian theories.  Although we have focused on four-dimensional examples, the connection between symmetry fractionalization and anomalies is more broad and can be explored in diverse spacetime dimensions. It has been particularly well studied in three dimensions, building on the systematic discussions in \cite{Barkeshli:2014cna, Barkeshli:2019vtb}.  Finally, it is interesting to extend the dictionary between the symmetry properties of line defects and the 't Hooft anomalies of the bulk quantum field theory to the situation where 1-form and 0-form symmetries are part of a non-trivial 2-group (or a more general higher group), with recent progress reported in~\cite{Barkeshli:2014cna, Bhardwaj:2021wif, Lee:2021crt, Apruzzi:2021mlh, Bhardwaj:2022dyt}.

\emph{Note Added:} While this paper was being completed we were informed of \cite{Delmastro:2022pfo} which also studies the relationship between symmetry fractionalization and anomalies.

\section{Line Defects and Anomalies in Maxwell Theory}
\label{maxlines}

In this section we explore the notion of line defect quantum numbers and symmetry fractionalization, and its relation to 't Hooft anomalies, in the simple context of Abelian gauge theories. Much of this section is a review and synthesis of previous discussions; see~\cite{Witten:1995gf, Thorngren:2014pza, Wang:2015fmi, Metlitski:2015yqa, Zou:2017ppq, Cordova:2018acb, Wang:2018qoy, Hsin:2019fhf, Ning:2019ffr, Ang:2019txy} for related work. We will apply the lessons learned here in section~\ref{rgflows}, to probe anomalies of non-Abelian gauge theories with matter by triggering suitable RG flows that connect them to Abelian gauge theories in the IR. 

\subsection{1-Form Symmetries and Wilson-'t Hooft Lines} 

Consider free Maxwell theory in four dimensions, i.e.~free~$U(1)$ gauge theory without matter. The gauge field~$a^{(1)}$ is a dynamical~$U(1)$ connection, with field strength~$f^{(2)} = da^{(1)}$. In addition to the bosonic~$SO(4)$ Euclidean spacetime symmetry, the only other connected and intrinsic symmetries of the theory are the electric and magnetic 1-form symmetries,\footnote{~The theory has a discrete charge-conjugation symmetry, and for certain values of the~$\theta$-angle, it also preserves parity and time reversal. We will not emphasize these discrete symmetries in the present paper, though they can be studied along similar lines.} 
\begin{equation}\label{eq:max1form}
G^{(1)} = U(1)_e^{(1)} \times U(1)_m^{(1)}~.
\end{equation}
Here we are following the terminology of~\cite{Gaiotto:2014kfa}, i.e.~an ordinary global symmetry that acts on local operators is a 0-form symmetry, while 1-form symmetries act on line defects. The symmetry $U(1)^{(1)}_{e}$ measures the electric charges of Wilson lines, while $U(1)_{m}^{(1)}$ measures the magnetic charges of 't Hooft lines.  The most general lines are dyons with quantized electric and magnetic charges~$(E, M) \in \IZ \times \IZ$. 

In slightly more detail, a Wilson line~$W_E(C)$ of charge~$E$ along a curve~$C$ is given by the holonomy of the gauge field along~$C$,
\begin{equation}
W_E(C) = \exp\left(i E \int_C a^{(1)}\right)~.
\end{equation}
Physically, these defects model the worldlines of infinitely heavy probe particles of electric charge~$E$.  

Meanwhile an 't Hooft line~$H_M(C)$ of charge~$M$ is defined as a disorder operator describing the worldline of an infinitely heavy Dirac monopole of magnetic charge~$M$. Explicitly, we can define~$H_M(C)$ by removing a small~$S^2_\varepsilon$ of radius~$\varepsilon \rightarrow 0$ around every point on the line and imposing boundary conditions for~$a^{(1)}$ on this sphere that fix the magnetic flux to be~${1 \over 2 \pi} \int_{S^2_\varepsilon} f^{(2)} = M$. 

A general dyonic line with charges~$(E, M)$ can be described as a (suitably regularized) operator product~$W_E(C) H_M(C)$. 

The existence of the 1-form symmetries~\eqref{eq:max1form} gives a clean way to identify the line defects in Maxwell theory as those carrying specific 1-form charges dictated by their electric and magnetic charges~$(E, M)$. The presence of a 1-form symmetry is the most robust way to identify lines in the presence of matter and interactions, because lines that are charged under such a symmetry cannot break or be screened away. Some of the examples discussed in section~\ref{rgflows} have a microscopic 1-form symmetry, while others do not, but in all examples much or all of the 1-form symmetry~\eqref{eq:max1form} of Maxwell theory emerges in the IR. 

\subsection{1-Form Symmetry Background Fields and Anomalies}
	
As usual, the~$U(1)_e^{(1)}$ and~$U(1)_m^{(1)}$ 1-form symmetries of Maxwell theory can be probed by coupling to suitable background gauge fields. Here they are 2-form background gauge fields~$B^{(2)}_e$ and~$B_m^{(2)}$, which couple to free Maxwell theory through the following action,\footnote{~For simplicity, we set the~$\theta$-angle of Maxwell theory to zero.}  
\begin{equation}\label{u1act}
S=\frac{1}{2g^{2}}\int_{M_4} \left(f^{(2)}-B_{e}^{(2)}
\right)\wedge *\left(f^{(2)}-B_{e}^{(2)}\right)+\frac{i}{2\pi}\int _{M_4} \left(f^{(2)}-B_{e}^{(2)}\right) \wedge B_{m}^{(2)}~.
\end{equation}
The transformation laws under background~$U(1)_e^{(1)} \times U(1)_m^{(1)}$ gauge transformations are as follows, \begin{equation}\label{gaugebb}
B_{e}^{(2)}\rightarrow B_{e}^{(2)}+d\lambda_{e}^{(1)}~, \hspace{.2in}B_{m}^{(2)}\rightarrow B_{m}^{(2)}+d\lambda_{m}^{(1)}~, \hspace{.2in} a^{(1)} \rightarrow a^{(1)}+\lambda_{e}^{(1)}~.
\end{equation}
The parameters $\lambda_{e,m}^{(1)}$ are locally 1-forms, but may themselves have non-vanishing quantized fluxes on closed 2-cycles~$\Sigma_2$, 
\begin{equation}
\frac{1}{2\pi}\int_{\Sigma_2} d\lambda_{e,m}^{(1)}\in \mathbb{Z}~.
\end{equation}
In other words~$\lambda_{e,m}^{(1)}$ are~$U(1)$ connections. The fact that~$a^{(1)}$ shifts inhomogenously indicates that it is a Nambu-Goldstone boson, and that the 1-form symmetry is spontaneously broken in Maxwell theory~\cite{Gaiotto:2014kfa}.

From the action \eqref{u1act} and the gauge transformations \eqref{gaugebb} we can conclude that the action is not exactly gauge invariant but rather transforms as follows,
\begin{equation}\label{ds}
S \rightarrow S -\frac{i}{2\pi}\int B_{e}^{(2)} \wedge d\lambda_{m}^{(1)}~.
\end{equation}
Since this shift only involves the background fields, it constitutes an 't Hooft anomaly of the $U(1)^{(1)}_{e}\times U(1)^{(1)}_{m}$ global symmetry~\cite{Gaiotto:2014kfa}. This anomaly may be cancelled (and hence characterized) via inflow from a five-dimensional mixed Chern-Simons theory for the background fields:
\begin{equation}\label{anomu1}
\mathcal{A}=\frac{i}{2\pi}\int_{M_{5}} dB_{e}^{(2)}\wedge B_{m}^{(2)}~.
\end{equation}
On closed five-manifolds~$M_{5}$ the action $\mathcal{A}$ yields a well-defined topological theory, but on manifolds with boundary~$\partial M_5 = M_4$ this action is not gauge invariant and yields (minus) the anomalous variation \eqref{ds} via inflow.  Similarly, on closed~$M_5$ the action~$\cal A$ treats the background fields $B_{e}^{(2)}$ and $B_{m}^{(2)}$ on equal footing, since we can integrate by parts. On manifolds with boundary $\partial M_5 = M_4$ this corresponds to modifying the theory on~$M_4$ by a local counterterm~$\sim \int_{M_4} B_e^{(2)} \wedge B_m^{(2)}$.

\subsection{Restriction to Discrete 1-Form Symmetry}

In the examples we analyze in section~\ref{rgflows}, only discrete subgroups of the 1-form symmetry are activated. We now explain how the background fields and the 't Hooft anomaly~\eqref{anomu1} specialize in this case. It will suffice to consider discrete subgroups of the form
\begin{equation}\label{subg}
\mathbb{Z}_{N,e}^{(1)}\times \mathbb{Z}_{N,m}^{(1)} \subset U(1)_{e}^{(1)}\times U(1)_{m}^{(1)}~.
\end{equation}
This restriction requires the background fields~$B_{e,m}^{(2)}$ to be flat,
\begin{equation}\label{flat}
dB_{e}^{(2)}=0~, \hspace{.2in}dB_{m}^{(2)}=0~.
\end{equation}
The gauge-invariant information in these flat background fields is then encoded in their~$\IZ_N$-valued holonomies, which define the following cohomology classes, 
\begin{equation}\label{smallbdef}
b_{e}^{(2)}\equiv \left[\frac{N}{2\pi}B_{e}^{(2)}\right]\in H^{2}(M_{4},\mathbb{Z}_{N})~, \hspace{.2in}b_{m}^{(2)}\equiv\left[\frac{N}{2\pi}B_{m}^{(2)}\right]\in H^{2}(M_{4},\mathbb{Z}_{N})~.
\end{equation}
Here~$[\cdots]$ denotes the gauge-equivalence class. With these restrictions on the background fields, the Maxwell action coupled to background fields takes the same form as in \eqref{u1act}.

Let us now turn to the 't Hooft anomaly~\eqref{anomu1}.  Superficially, it vanishes because the background fields satisfy the flatness condition~\eqref{flat}, but these conditions only hold at the level of differential forms and hence they are blind to torsion. To extract this additional information, we make use of the Bockstein homomorphism~$\beta_N$ associated with the short exact sequence~$0 \longrightarrow \IZ_N \overset{\times N} \longrightarrow \IZ_{N^2} \longrightarrow \IZ_N \longrightarrow 0$, where the second arrow is multiplication by~$N$. Given~$X_2 \in H^2(M_4, \IZ_N)$, the Bockstein homomorphism defines a cohomology class~$\beta_N(X_2) \in H^3(M_4, \IZ_N)$ via
\begin{equation}\label{bockdef}
\beta_N(X_2) \equiv \frac{1}{N} \delta \t X_2  \in H^{3}(M_{4},\mathbb{Z}_N)~.
\end{equation}
Here~$\t X_2$ is any choice of integral co-chain that reduces to~$X_2$ modulo~$N$.\footnote{~Since~$X_2$ is a~$\IZ_N$ class, $\delta \t X_2$ vanishes modulo~$N$, and hence~${1 \over N} \delta \t X_2$ is a well-defined integer co-chain. It is clearly closed, and picking different lifts~$\t X_2$ changes it by an exact co-chain, so that it defines a cohomology class in~$H^3(M_4, \IZ)$ that only depends on~$X_2$. Since~$N \beta_N(X_2)$ vanishes in cohomology, it follows that~$\beta_N(X_2)$ is always~$N$-torsion. } The Bockstein homomorphism~$\beta_N$ is part of a long exact sequence in cohomology,
\begin{equation}
\cdots \longrightarrow H^2(M_4, \IZ_{N^2}) \longrightarrow H^2(M_4, \IZ_N) \overset{\beta_N} \longrightarrow H^3(M_4, \IZ_N) \longrightarrow \cdots~,
\end{equation}
which shows that~$\beta_N(X_2)$ is the obstruction to lifting~$X_2$ from~$H^2(M_4, \IZ_N)$ to~$H^2(M_4, \IZ_{N^2})$. 

It can be seen from the definition~\eqref{bockdef} of the Bockstein homomorphism that~$\beta_N(b_e^{(2)})$ is a discrete version of the field strength~${1 \over 2 \pi} dB_e^{(2)}$. This observation allows us to reduce the 't Hooft anomaly in~\eqref{anomu1} from~$U(1)_e^{(1)} \times U(1)_m^{(1)}$ to~$\mathbb{Z}_{N,e}^{(1)}\times \mathbb{Z}_{N,m}^{(1)}$,
\begin{equation}\label{discreteanom}
\mathcal{A}=\frac{2\pi i}{N}\int_{M_{5}} \beta_N\left(b_{e}^{(2)}\right)\cup b_{m}^{(2)}=\frac{2\pi i}{N}\int_{M_{5}} b_{e}^{(2)}\cup \beta_N \left(b_{m}^{(2)}\right)~.
\end{equation}
Here we have used the fact that~$\beta_N$ obeys a Leibniz rule on closed manifolds~$M_5$, so that it can be integrated by parts.

\subsection{Enriching by 0-Form Global Symmetry}

We now describe how Maxwell theory can be enriched by an ordinary (i.e.~0-form) global symmetry~$G^{(0)}$.  For reasons explained below, we focus on connected (and hence continuous)~$G^{(0)}$. Our primary interest is in the quantum numbers of line defects with respect to this symmetry group.  For now, we focus on the case of internal symmetries; the generalization to spacetime symmetries is discussed in section~\ref{bflines}.

Many aspects of QFT with~$G^{(0)}$ symmetry are controlled by the Lie algebra of~$G^{(0)}$, i.e.~the local form of~$G^{(0)}$. For instance, it is this local form that controls the conserved currents associated to the symmetry. By contrast, our discussion will only be sensitive to the global form of~$G^{(0)}$, and local aspects will essentially not play any role. We must therefore carefully distinguish between~$G^{(0)}$ and various possible discrete quotients or extensions that naturally arise, e.g.~we must distinguish between systems with~$SU(2)$ and~$SO(3)$ global symmetry. 

Mathematically, the symmetry group $G^{(0)}$ fits into the following short exact sequence of groups,
\begin{equation}
\label{seq}
1~ \longrightarrow~ Z^{(0)} ~\longrightarrow~ \widetilde{G}^{(0)} ~ \longrightarrow ~G^{(0)} ~\longrightarrow~ 1~.
\end{equation}
Here, $\widetilde{G}^{(0)}$ is the simply-connected universal cover of the group $G^{(0)}$, while $Z^{(0)}$ is a (necessarily abelian) subgroup of the center of $\widetilde{G}^{(0)}$.  Thus $ G^{(0)}\cong \widetilde{G}^{(0)}/Z^{(0)}$ and has fundamental group $\pi_{1}(G^{(0)})\cong Z^{(0)}$. In the examples we study, $G^{(0)}$ will be a non-Abelian semisimple Lie group and~$Z^{(0)}$ will be a finite abelian group.  An instructive example is~$G^{(0)}  = SO(3)$, for which the sequence~\eqref{seq} takes the form
\begin{equation}
1~ \longrightarrow ~ Z^{(0)} = \mathbb{Z}_{2}^{(0)} ~ \longrightarrow ~ \widetilde{G}^{(0)}  \cong SU(2)~ \longrightarrow ~ G^{(0)} \cong SO(3) ~ \longrightarrow 1~.
\end{equation}

One way to access the global observables that are sensitive to the quotient by~$Z^{(0)}$ is to couple the theory to background gauge fields defining a~$G^{(0)}$ bundle.  Not all such bundles can be lifted to bundles of the simply connected covering group~$\widetilde{G}^{(0)}$, and the class that measures this obstruction is a (generalized) Stiefel-Whitney class
\begin{equation}\label{w2def}
w_{2}(G^{(0)})\in H^{2}(M_{4}, Z^{(0)})~,
\end{equation}
i.e.~this class vanishes if and only if the~$G^{(0)}$ bundle can be lifted to a~$\widetilde{G}^{(0)}$ bundle. Conversely, the ability to consider the theory in the presence of backgrounds with non-zero~$w_{2}(G^{(0)})$ encodes the global form of~$G^{(0)}$. Physically, we can think of this class as defining a discrete magnetic flux for the~$G^{(0)}$ background gauge fields. 

\subsubsection{0-Form Symmetry Fractionalization}

Since we are focusing on connected~$G^{(0)}$, no local operators in Maxwell theory transform under the~$G^{(0)}$ symmetry.\footnote{~Maxwell theory has discrete charge-conjugation and (for certain~$\theta$-angles) parity and time-reversal symmetries, all of which act on local operators.} This, however, does not imply that Maxwell theory cannot be coupled to the~$G^{(0)}$ background fields in an interesting way. Indeed, as we now review, $G^{(0)}$ can act on the line defects of the theory through a mechanism known as symmetry fractionalization, which is captured by precisely such a coupling.  

Intuitively, the action of~$G^{(0)}$ on line defects can be understood via the action of the symmetry on heavy probe particles represented by those lines. Since these particles can transform under~$G^{(0)}$, so should the lines. While this intuition is in essence correct, it is important to note that most of this information is scheme dependent; equivalently, it is UV sensitive and cannot be unambiguously encoded in Maxwell theory itself. 

This scheme dependence is described by an analogue of local operator wavefunction renormalization for line defects. Recall that the latter involves rescaling a local operator~$\CO(x) \rightarrow K \CO(x) $ by a scheme-dependent constant~$K$. The correct generalization to a line defect~$L(C)$ extended along~$C$ is to allow rescalings by local functionals of the background fields along~$C$. A standard example involves rescaling~$L(C)$ by a cosmological constant counterterm on the line,\footnote{~Here~$M$ can be thought of as a mass (or mass shift) for the heavy probe particle represented by the line.} 
\begin{equation}
L(C) \quad \longrightarrow \quad L(C) \, \exp \left(M \int_{C} ds\right)~.
\end{equation}

In our context, we are interested in counterterms constructed using the background gauge fields~$A^{(1)}$ for the~$G^{(0)}$ symmetry. These are Wilson lines for~$A^{(1)}$, in genuine representations ${\bf R} \in \text{Rep}(G^{(0)})$. Dressing a line defect~$L(C)$ by such a Wilson line counterterm
\begin{equation}\label{eq:linect}
L(C) \quad \longrightarrow \quad L(C) \Tr_{\bf R} \left( \exp \left(i \int_C A^{(1)} \right)\right)~,
\end{equation}
indicates that the line transforms in the representation~${\bf R}$ of~$G^{(0)}$. We conclude that this representation data is scheme dependent.

After modding out by this scheme dependence, we can ask whether there is any remaining physical (i.e.~scheme-independent) sense in which we can assign~$G^{(0)}$ quantum numbers to the lines. Thankfully, the answer is affirmative (modulo some assumptions, see below), because the lines may transform in projective rather than genuine representations of~$G^{(0)}$. Recall that a projective representation of~$G^{(0)}$ is one in which the group multiplication law only holds up to phases. Equivalently, a projective representation of~$G^{(0)}$ is simply an ordinary representation of the universal covering group~${\widetilde G}^{(0)}$. One intuitive way to see why the lines may transform in projective representations of~$G^{(0)}$ is that they represent heavy particles that carry gauge as well as global charges. While the gauge quotient reduces the symmetry to~$G^{(0)}$ on gauge-invariant local operators, additional symmetry can be liberated on line defects. 

At this point we pause to mention two phenomena that obstruct our ability to assign projective representations to line defects:
\begin{itemize}
\item In general a disconnected 0-form symmetry can permute line defects by acting on their electric or magnetic charges. Only line defects that are not permuted can be assigned a consistent projective representation under such a symmetry. Since we are only considering connected symmetries~$G^{(0)}$, this issue does not arise.

\item In general the~$G^{(0)}$ symmetry may be part of a non-trivial 2-group (or an even higher group) together with the electric and magnetic 1-form symmetry~$G^{(1)} = U(1)_e^{(1)} \times U(1)_m^{(1)}$. (See for instance~\cite{Tachikawa:2017gyf,Cordova:2018cvg,Benini:2018reh}.) If this 2-group has non-vanishing Postnikov class in~$H^3(G^{(0)}, G^{(1)})$, then it is not possible to consistently specify projective~$G^{(0)}$ representations for certain lines charged under~$G^{(1)}$. For this reason such a class is sometimes referred to as an obstruction to~$G^{(0)}$ symmetry fractionalization~\cite{Barkeshli:2014cna}. Going forward, we explicitly assume that~$G^{(0)}$ is not part of such a 2-group. 
\end{itemize}

Even in the absence of these obstructions, it does not make sense to ask exactly which projective representation of~$G^{(0)}$ a line transforms in. This is because our discussion of line counterterms around~\eqref{eq:linect} implies that tensoring with genuine representations of~$G^{(0)}$ is scheme dependent. We can therefore only unambiguously determine the equivalence class of the projective representation modulo ordinary ones. In the example~$G^{(0)} = SO(3)$ this means that it makes sense to ask whether a line has integer~${\widetilde G}^{(0)} = SU(2)$ spin (genuine) or half-integer~${\widetilde G}^{(0)} = SU(2)$ spin (projective), but the precise value of the spin is scheme dependent. 

More generally, consider any irreducible representation $\mathbf{R} \in \text{Rep}(\widetilde{G}^{(0)})$, or equivalently a projective representation of~$G^{(0)}$. Then in $\mathbf{R}$, the central group $Z^{(0)}$ in \eqref{seq} that defines the quotient~$G^{(0)} = {\widetilde G}^{(0)} / Z^{(0)}$ acts by phases.  To make this explicit, we write~$Z^{(0)}$ as a product of cyclic groups,
\begin{equation}\label{zdecomp}
Z^{(0)} \cong \mathbb{Z}_{n_{1}}\times \mathbb{Z}_{n_{2}}\times \cdots \times \mathbb{Z}_{n_{k}}~.
\end{equation}
A representation of~$Z^{(0)}$ is then characterized by a set of~$k$ discrete charges, which we parametrize via the following~$k$-tuple of fractions,
\begin{equation}\label{fracdef}
\vec Q\equiv \left(\frac{q_{1}}{n_{1}},\frac{q_{2}}{n_{2}},  \cdots, \frac{q_{k}}{n_{k}}\right)~, \hspace{.2in} q_{i}\sim q_{i}+n_{i} \quad (i = 1 , \ldots, k) ~.
\end{equation}
A general group element $\vec z=(\ell_{1}, \ldots , \ell_{k})\in Z^{(0)}$ with~$\ell_i \in \IZ_{n_i}$ then acts on vectors in the representation $\mathbf{R}$ as follows,
\begin{equation}
\vec z\mapsto \exp\left(2\pi i \vec Q\cdot \vec z\right)=\exp\left(2\pi i\left(\frac{\ell_{1}q_{1}}{n_{1}}+\frac{\ell_{2}q_{2}}{n_{2}}+\cdots +\frac{\ell_{k}q_{k}}{n_{k}}\right)\right)~.
\end{equation}
Tensoring with any genuine representation of $G^{(0)}$ cannot modify the charge vector $\vec Q$ since, by definition, $Z^{(0)}$ acts trivially in such a representation. Thus all scheme-independent information about the projective~$G^{(0)}$ action on the lines is contained in the fractional charge vector~$\vec Q$. 

The phenomenon of line defects transforming in projective representations of~$G^{(0)}$ is often referred to as symmetry fractionalization. The fractions~${q_i / n_i}$ that define the charge vector~$\vec Q$ encode the discrepancy between the integer charges of local operators in genuine representations of~$G^{(0)}$ and those of the projectively transforming lines.

\subsubsection{0-Form Symmetry Background Fields and Anomalies}

We now describe how to couple Maxwell theory to~$G^{(0)}$ background fields to describe the fractionalized~$G^{(0)}$ quantum numbers of the lines. It suffices to characterize the~$G^{(0)}$ quantum numbers of the fundamental Wilson line~$W_{E = 1}(C)$ of electric charge~$E = 1$ and the fundamental 't Hooft line~$H_{M = 1}(C)$ of magnetic charge~$M = 1$. The behavior of all other lines is fixed by fusion and~$\mathsf{ CPT}$ symmetry.  

To this end, recall that quantum mechanics (i.e.~one-dimensional QFT) with global symmetry~$G^{(0)}$ has a possible 't Hooft anomaly classified by projective representations of $G^{(0)}$ modulo equivalence under tensoring with genuine representations of $G^{(0)}$. These anomalies are therefore labeled by exactly the same data as the vector~$\vec Q$ of fractional charges in~\eqref{fracdef} above.  Physically, the meaning of this anomaly is that the operators in the quantum mechanics transform in genuine representations of $G^{(0)}$ while the states all transform in projective representations characterized by~$\vec Q$. By acting with operators on states in the Hilbert space, we tensor these states with genuine representations of~$G^{(0)}$ leading to different projective representations, but the projective equivalence class (i.e.~the representation of~$Z^{(0)} \subset {\widetilde G}^{(0)}$) defined by~$\vec Q$ remains inert.  In our problem we have an entirely analogous situation, except that the quantum mechanics in question is the worldvolume theory of a line defect.

The quantum mechanical anomaly described above can be described by inflow from a two-dimensional classical action for the~$G^{(0)}$ background fields that is constructed using the Stiefel-Whitney class discussed around \eqref{w2def}, which we repeat here,
\begin{equation}\label{w2defrep}
w_{2}(G^{(0)})\in H^{2}(M_{4}, Z^{(0)})~.
\end{equation}
Recalling the decomposition~$Z^{(0)} \cong \mathbb{Z}_{n_{1}}\times \mathbb{Z}_{n_{2}}\times \cdots \mathbb{Z}_{n_{k}}$ in \eqref{zdecomp}, one has an associated decomposition of the Stiefel-Whitney class, 
\begin{equation}\label{wdecomp}
\vec w_{2}(G^{(0)})=\left(w_{2,1}(G^{(0)}), w_{2,2}(G^{(0)}), \cdots, w_{2,k}(G^{(0)})\right)~, \hspace{.2in}w_{2,i}(G^{(0)}) \in H^{2}(M_{4},\mathbb{Z}_{n_{i}})~.
\end{equation}
The anomaly implies that in the presence of general~$G^{(0)}$ background fields, a line~$L(C)$ transforming projectively under~$G^{(0)}$, characterized by a fractional charge vector~$\vec Q$ (see~\eqref{fracdef}), can be described as the boundary~$C = \partial \Sigma_2$ of a two-dimensional open surface~$\Sigma_2$ that supports the following anomaly inflow action,
\begin{equation}\label{surfSPT}
S=2\pi i\int_{\Sigma_2} \vec Q\cdot \vec w_{2}(G^{(0)})=2\pi i\int_{\Sigma_2 }\left(\frac{q_{1}}{n_{1}}w_{2,1}(G^{(0)})+\frac{q_{2}}{n_{2}}w_{2,2}(G^{(0)})+\cdots +  \frac{q_{k}}{n_{k}}w_{2,k}(G^{(0)})\right)~.
\end{equation}
Note that the dependence on~$\Sigma_2$ is topological.  

The worldline anomaly action \eqref{surfSPT} describes how a line bounding~$\Sigma_2$ couples to background $G^{(0)}$ fields.  We must now upgrade this to a bulk coupling of Maxwell theory to $G^{(0)}$ background fields such that \eqref{surfSPT} is correctly induced on lines.  This can be accomplished using the 1-form symmetries of Maxwell theory.  

Consider a dyonic line $L(C)$ with integer electric and magnetic charges $(E,M) \in \IZ \times \IZ$. Since~$(E, M)$ are simultaneously the charges under the~$U(1)_e^{(1)} \times U(1)_m^{(1)}$ global symmetry of Maxwell theory, they can be detected by topological surface operators supported on an~$S^2$ linking the support~$C$ of the line in spacetime, 
\begin{equation}\label{EMcharges}
E = {1 \over g^2} \int_{S^{2}} * f^{(2)}~, \qquad M = {1 \over 2 \pi} \int_{S^{2}} f^{(2)}~.
\end{equation}
Comparing with the Maxwell action~\eqref{u1act}, we see that turning on flat~$U(1)_e^{(1)} \times U(1)_m^{(1)}$ background fields~$B_{e,m}^{(2)}$, or their specialization to~$b_{e,m}^{(2)} \in H^2(M_4, \IZ_N)$ in~\eqref{smallbdef}, attaches the following two-dimensional surface integral to the line,
\begin{equation}\label{battachline}
L(C) \longrightarrow L(C) \, \exp\left(i  \int_{\Sigma_2} \left(E B_e^{(2)} + M B_m^{(2)} \right) \right) = L(C) \, \exp\left({2 \pi i  \over N} \int_{\Sigma_2} \left( E b_e^{(2)} + M  b_m^{(2)}\right) \right)~,
\end{equation}
with~$C = \partial \Sigma_2$. This is because the integral over the spacetime manifold~$M_4$ locally factorizes into an integral over the~$S^2$ linking~$C$ and the surface~$\Sigma_2$ bounded by it.

Let~$\vec Q_e$ be the fractionalized~$G^{(0)}$ charge vector of the charge-one Wilson line $W_1(C)$, and let~$\vec Q_m$ be the corresponding charge vector of the charge-one 't Hooft line~$H_1(C)$. By fusion, the charge vector of the dyonic~$(E, M)$ line is~$\vec Q = E \vec Q_e + M \vec Q_m$. Comparing the anomaly inflow action~\eqref{surfSPT} with~\eqref{battachline}, we deduce the required 1-form background fields, 
\begin{equation}\label{backgroundid}
\frac{b_{e}^{(2)}}{N} = \vec Q_{e} \cdot \vec w_{2}(G^{(0)})~, \qquad \frac{b_{m}^{(2)}}{N} = \vec Q_{m} \cdot \vec w_{2}(G^{(0)})~.
\end{equation}
In this formula, we have taken~$N=\text{lcm}(n_{1}, n_{2}, \cdots, n_{k})$ and~$b_{e,m}^{(2)} \in H^2(M_4, \IZ_N)$. 	However, recall from~\eqref{wdecomp} that the vector~$\vec w_2(G^{(0)})$ is comprised of cohomology classes~$w_{2, i}(G^{(0)}) \in H^2(M_4, \IZ_{n_i})$. In~\eqref{backgroundid}, we have implicitly mapped these into~$H^2(M_4, \IZ_N)$ using the short exact sequence~$0 \rightarrow \IZ_{n_i} \rightarrow \IZ_{N} \rightarrow \IZ_{N/n_i} \rightarrow 0$, which implies the following long exact sequence,\footnote{~Note that the map~$H^2(M_4, \IZ_{n_i}) \rightarrow H^2(M_4, \IZ_N)$ has a kernel given by the image of~$H^1(M_4, \IZ_{N/n_i})$ under the Bockstein map. This kernel vanishes if we restrict to simply-connected spacetimes~$M_4$, which are sufficient to diagnose all phenomena we study here. } 
\begin{equation}\label{longexii}
\cdots \rightarrow H^1(M_4, \IZ_{N/n_i}) \rightarrow H^2(M_4, \IZ_{n_i}) \rightarrow H^2(M_4, \IZ_N)\rightarrow \cdots~.
\end{equation} 
In practice, this simply means writing the coefficients~$q_i/n_i$ of all~$w_{2, i}(G^{(0)})$ in~\eqref{backgroundid} over the common denominator~$N$. Crucially, it does not imply that~$w_{2, i}(G^{(0)})$ is a reduction modulo~$n_i$ of a~$\IZ_N$ class (the existence of such a class is generally obstructed). 

To deduce the 't Hooft anomaly of $G^{(0)}$-enriched Maxwell theory, we substitute the background fields~\eqref{backgroundid} into~\eqref{discreteanom},
\begin{equation}\label{anomform}
\mathcal{A}=2\pi i N \int \beta_N\left(\vec Q_{e}\cdot \vec w_{2}(G^{(0)})\right) \cup \left(\vec Q_{m}\cdot \vec w_{2}(G^{(0)})\right) = 2\pi i N \int \left(\vec Q_{e}\cdot \vec w_{2}(G^{(0)})\right) \cup \beta_N \left( \vec Q_{m}\cdot \vec w_{2}(G^{(0)})\right)~.
\end{equation}
Note that this anomaly is completely dictated by the fractionalized~$G^{(0)}$ charges~$\vec Q_{e, m}$ of the fundamental Wilson and 't Hooft lines. In section~\ref{rgflows} we will use this fact to detect 't Hooft anomalies in non-Abelian gauge theories by flowing to pure Maxwell theory and examining the pattern of symmetry fractionalization.

\subsection{Bosonic and Fermionic Lines}\label{bflines}

In the preceding subsection we assumed that~$G^{(0)}$ was a connected internal symmetry. Here we generalize this to spacetime symmetries by including in~$G^{(0)}$ the connected part of the Lorentz group. 

As was the case for internal symmetries, we must first determine the global form of the spacetime symmetry. Here we must distinguish between a bosonic (or non-spin) QFT, where all local operators are bosons, and a fermionic (or spin) QFT where some local operators are fermions. In the fermionic case the Euclidean spacetime symmetry is~$\text{Spin}(4)$, and the theory can only be formulated on a spin manifold, for which the second Stiefel-Whitney class of the tangent bundle vanishes, $w_2(M_4) = 0$. By contrast, the spacetime symmetry in a bosonic theory is~$SO(4)$, and the theory can be studied on manifolds with~$w_2(M_4) \neq 0$.\footnote{~We continue to assume that~$M_4$ is oriented, so that~$w_1(M_4) = 0$.} We thus conclude that
\begin{equation}
G^{(0)} \supset \text{Spin}(4) \quad (\text{fermionic QFT})~, \qquad G^{(0)} \supset SO(4) \quad (\text{bosonic QFT})~.
\end{equation}

Now let us turn to the quantum numbers of line defects. If the theory is fermionic, there is no candidate fractionalization class in~$H^2(M_4, \IZ)$. Relatedly, $\text{Spin}(4)$ is simply connected and has no projective representations. Indeed, any given line~$L(C)$ may be dressed by a background Wilson line counterterm of the form~\eqref{eq:linect}, with the background gauge field~$A^{(1)}$ replaced by the spin connection~$\omega^{(1)}$. This line can be in any representation~${\bf R} \in \text{Rep}\left(\text{Spin}(4)\right)$, so that neither the spin nor the statistics of the line~$L(C)$ is scheme-independent. 

By contrast, in bosonic theories the only available counterterms are for those~${\bf R}$ which are genuine representations of~$SO(4)$. We can therefore distinguish between bosonic and fermionic lines, which transform trivially and non-trivially under the fermion parity~$(-1)^F$ that extends~$SO(4)$ to~$\text{Spin}(4)$. As was argued in~\cite{Witten:1995gf,Thorngren:2014pza} (see for instance ~\cite{Cordova:2018acb, Wang:2018qoy} for a review), the two-dimensional anomaly inflow action that should be attached to a fermionic line~$L(C)$ in a bosonic QFT is
\begin{equation}\label{tangent}
\exp\left(i \pi  \int_{\Sigma_2}w_{2}(M)\right)~, \qquad \partial \Sigma_2 = C~.
\end{equation}
The choice of~$\Sigma_2$ is topological; it is somewhat analogous to choosing a branch cut for fermionic point operators in two dimensions. 

Let us sketch how the inflow action \eqref{tangent} can be used to argue that the line~$L(C)$ is a fermion. If~$M_4$ is a spin manifold, then~$w_2(M_4)$ vanishes in cohomology. We can thus write~$w_2(M_4) = \delta \gamma_1$, where~$\gamma_1 \in C^1(M_4, \IZ_2)$ is a co-chain with~$\IZ_2$ coefficients. In fact, since~$\gamma_1$ trivializes~$w_2(M_4)$, it defines a spin structure on~$M_4$. The inflow action~\eqref{tangent} then reduces to a local counterterm along the support~$C$ of the line, which encodes the expected coupling of a fermion to the spin structure on a spin manifold, 
\begin{equation}\label{tangent1}
\exp\left(i \pi \int_{\Sigma_2}w_{2}(M_4)\right) = \exp\left(i \pi  \int_{C}\gamma_1 \right)~.
\end{equation}
We can apply the same logic to the inflow action~\eqref{surfSPT} for~$G^{(0)}$ flavor symmetries. If we restrict to bundles of the simply connected~${\widetilde G}^{(0)}$ symmetry,  all the Stiefel-Whitney classes in~\eqref{surfSPT} can be trivialized, and the anomaly inflow action can be written as a background Wilson line for~${\widetilde G}^{(0)}$, in a representation that is fixed by the fractional charge vector~$\vec Q$ (modulo genuine~$G^{(0)}$ representations).

\subsubsection{Background Fields and $\text{Spin}_{c}$ Connections}\label{spincsec}

Since Maxwell theory is bosonic, we can generalize our previous analysis of charge fractionalization to include the statistics of lines.  We add to $b_{e}^{(2)}$ and $b_{m}^{(2)}$ contributions involving $w_{2}(M_4)$,
\begin{equation}\label{sigdef}
\sigma_{e} w_{2}(M_4 )\subset b_{e}^{(2)}~, \hspace{.5in}\sigma_{m} w_{2}(M_4 )\subset b_{m}^{(2)}~,
\end{equation}
where we have introduced $\mathbb{Z}_{2}$-valued variables $\sigma_{e,m}$ that encode the statistics of the charge-one Wilson and 't Hooft lines~$W_1$ and~$H_1$ as follows, 
\begin{equation}\label{sigmadef}
\sigma_{e}=\begin{cases} 0 &W_1 \text{ line bosonic}~,\\ 1 & W_1 \text{ line fermionic}~, \end{cases}\hspace{.2in}\sigma_{m}=\begin{cases} 0 & H_1 \text{ line bosonic}~,\\
1  & H_1 \text{ line fermionic}~.
\end{cases}
\end{equation}

In general, $G^{(0)}$ will contain both internal flavor symmetries as well as the spacetime symmetry. In this case, we modify our fractional charge vectors $\vec Q_{e,m}$ in~\eqref{fracdef} and Stiefel-Whitney classes in~\eqref{wdecomp} to also include contributions from the statistics,
\begin{equation}\label{flavortospin}
\vec Q_{e}\rightarrow \left(\vec Q_{e, \text{flavor}}, \frac{\sigma_{e}}{2}\right)~, \hspace{.2in}\vec Q_{m}\rightarrow \left(\vec Q_{m, \text{flavor}}, \frac{\sigma_{m}}{2}\right)~,\hspace{.2in}\vec w_{2}(G^{(0)}) \rightarrow (\vec w_{2, \text{flavor}} , w_{2}(M_4))~.
\end{equation}
With these modifications, the required electric and magnetic 2-form background gauge fields needed to implement the fractionalization then take exactly the same form as in~\eqref{backgroundid}, 
\begin{equation}\label{backgroundid2}
\frac{b_{e}^{(2)}}{N}=\vec Q_{e} \cdot \vec w_{2}(G^{(0)})~, \hspace{.2in}\frac{b_{m}^{(2)}}{N}=\vec Q_{m} \cdot \vec w_{2}(G^{(0)})~,
\end{equation}
except that now $N=\text{lcm}(n_{1}, n_{2}, \cdots, n_{k},2)$.

Let us mention an alternative way to capture the fractionalized quantum numbers of the Wilson line: the Maxwell action \eqref{u1act} only depends on the combination $f^{(2)}-B_{e}^{(2)}$.  In the presence of the discrete backgrounds defined above, it is sometimes convenient to define a new dynamical gauge field $\tilde{a}^{(1)}$, whose curvature satisfies
\begin{equation}\label{twisting}
\frac{d\tilde{a}^{(1)}}{2\pi}=\frac{f^{(2)}}{2\pi}-\frac{B_{e}^{(2)}}{2\pi}=\frac{da}{2\pi}-\frac{b_{e}^{(2)}}{N}=\frac{da^{(1)}}{2\pi}-\left(\frac{q_{1,e}}{n_{1}}w_{2,1}+\frac{q_{2,e}}{n_{2}}w_{2,2}+\cdots + \frac{q_{k,e}}{n_{k}}w_{2,k}+\frac{\sigma_{e}}{2}w_{2}(M_4)\right)~.
\end{equation}
Locally, $\tilde{a}^{(1)}$ is a 1-form gauge field, but globally it is twisted: the periods of $\frac{d\tilde{a}}{2\pi}$ are fractions dictated by the Stiefel-Whitney classes on right-hand-side of \eqref{twisting}.

An example of such a twisted gauge field $\tilde{a}$ that is frequently discussed in the literature is a $\text{spin}_{c}$ connection. This corresponds to setting all~$q_{i, e} = 0$ in~\eqref{twisting}, as well as setting~$\sigma_e = 1$, so that 
\begin{equation}\label{spincab}
\int_{\Sigma_2} \frac{d\tilde{a}^{(1)}}{2\pi} \in \frac{1}{2}\int_{\Sigma_2}  w_{2}(M_4) + \IZ~.
\end{equation}
Similarly, a theory where the 't Hooft line is a fermion (with $\sigma_m = 1$) can be described by saying that the~$S$-dual electromagnetic gauge field is a $\text{spin}_{c}$ connection.

\subsubsection{Gravitational Anomalies and All-Fermion Electrodynamics}\label{sec:allfed}

The anomaly discussion around \eqref{anomform} applies straightforwardly to situations where the lines carry well-defined Bose or Fermi statistics. This leads to various possible 't Hooft anomalies involving internal and spacetime symmetries.  

Let us pause to mention an anomaly of this type that is purely gravitational, i.e.~it does not involve the internal symmetries, which is captured by the following inflow action, \begin{equation}\label{gravanom}
\mathcal{A}= i \pi \int_{M_5}  w_{2}(M_5)\cup \beta_2(w_{2}(M_5 ))=i \pi \int_{M_5} w_{2}(M_5)\cup w_{3}(M_5)~.
\end{equation} 
Here $ \beta_2(w_{2}(M_5))=w_{3}(M_5)$ is the third Stiefel-Whitney class of the spacetime manifold.  This $\mathbb{Z}_{2}$-valued anomaly is the only purely gravitational anomaly in four dimensions, see for instance~\cite{Kapustin:2014tfa, Thorngren:2014pza, Freed:2016rqq}. The action \eqref{gravanom} is non-vanishing on the unique non-trivial five-dimensional bordism class defined by the mapping torus of complex conjugation acting on $\mathbb{CP}^{2}$.  This implies that this gravitational anomaly can be detected by placing a QFT on $\mathbb{CP}^{2}$ and checking whether the partition function is even or odd under the large diffeomorphism defined by complex conjugation of the $\mathbb{CP}^{2}$ homogeneous coordinates. (See for instance~\cite{Wang:2018qoy} where this analysis is explicitly carried out in some examples.) 

An alternative way to understand this anomaly is via the statistics of lines, as discussed for instance in~\cite{Thorngren:2014pza, Kravec:2014aza,  Wang:2018qoy}. Consider the basic charge-one Wilson and 't Hooft lines~$W_1$ and~$H_1$, as well as the dyonic line~$D_{(1,1)} = W_1 H_1$ of charge~$(E, M) = (1,1)$ obtained by fusing them. Note that this fusion process imparts an additional half-unit of angular momentum~$J$ onto the dyon, which arises from the non-vanishing Dirac pairing between their charges,
\begin{equation}
J=\frac{1}{2} E M = \half~.
\end{equation}
This in turn means that the statistics label~$\sigma_d$ of the dyonic~$D_{(1,1)}$ line (defined in analogy with the statistics labels~$\sigma_{e,m}$ in~\eqref{sigmadef}) is shifted with respect to the naive sum~$\sigma_e + \sigma_m$ of the statistics labels of the Wilson and 't Hooft lines, 
\begin{equation}\label{spinfusion}
\sigma_{e}+\sigma_{m}+\sigma_{d} \equiv 1 \; (\text{mod }2)~.
\end{equation}
 
Note that~\eqref{spinfusion} has two classes of solutions.  The most familiar solutions are those where any two lines are bosonic, while the third line is fermionic.  This includes the case where the Wilson line and 't Hooft line are bosonic. Unless stated otherwise, this assignment of statistics is often implicit in discussions of Maxwell theory. Note that this choice is not invariant under electric-magnetic duality.  Specifically, the element $ST\in SL(2,\mathbb{Z})$ (which obeys $(ST)^{3}=1$) permutes the three possible choices of statistics described above (see for instance~\cite{Metlitski:2015yqa}). Since the coefficient of gravitational anomaly~\eqref{gravanom} is proportional to~$\sigma_e \sigma_m$, the anomaly vanishes for these choices of line statistics.  

The second class of solution to \eqref{spinfusion} is~$\sigma_e = \sigma_m = \sigma_d = 1$, so that all three lines are fermionic. This choice is $SL(2,\mathbb{Z})$ invariant.  The resulting version of Maxwell theory is sometimes referred to as all-fermion electrodynamics (see for instance \cite{Thorngren:2014pza,Kravec:2014aza,Wang:2018qoy}). Now~$\sigma_e \sigma_m = 1$ and the gravitational anomaly does indeed reproduce~\eqref{gravanom}.

\section{Anomalies from RG Flows into Maxwell Theory}
\label{rgflows}

In this section we discuss non-Abelian gauge theories that can flow to Maxwell theory at long distances.  By tracking the quantum numbers of lines in the IR, we will deduce the anomalies of the UV theory. We explicitly carry out this analysis for the simplest possible setting, namely that of~$SU(2)$ gauge theories.

All examples we consider have~$N_f$ left-handed Weyl fermions, schematically denoted by~$\Psi$, each of which transforms in the same irreducible representation $\mathbf{D}$ (with dimension $D = |{\bf D}|$) of the~$SU(2)$ gauge group.\footnote{~For sufficiently large $N_{f}$ or $\mathbf{D}$, this model is not asymptotically free.  In this case we view the theory as an effective QFT defined at some scale.  This in no way modifies the anomaly analysis to follow.}  Additionally we also include in our model a single real scalar field~$\Phi$ in the adjoint representation of $SU(2)$.  The Lagrangian takes the following schematic form,
\begin{equation}\label{nonabactiondef}
{\mathscr L}=  \Tr\Big\{{1 \over  g^2} F^2+(D \Phi)^2\Big\}+i \bar\Psi \slashed{D} \Psi+V(\Phi)+\left( {\mathscr L}_\text{Yukawa}+(\text{h.c.})\right)~. 
\end{equation}
Here~$F$ is the~$SU(2)$ field strength, $V(\Phi)$ is a scalar potential and ${\mathscr L}_\text{Yukawa} \sim \Phi \Psi \Psi$ is a Yukawa interaction.  We assume that $V(\Phi)$ is chosen so that the adjoint scalar $\Phi$ condenses and Higgses the $SU(2)$ gauge group down to $U(1)$. We also assume that the Yukawa couplings can be chosen in such a way that all fermions acquire a mass upon Higgsing.\footnote{~As we will see, this places additional constraints on~$N_f$ and~$\bf D$, beyond those that follow from gauge-anomaly cancellation.} Finally, by tuning the phase of the Yukawa couplings we can ensure that the~$\theta$-angle in the IR Maxwell theory vanishes,
\begin{equation}\label{irthetazero}
\theta_\text{Maxwell}=0~,
\end{equation}
as we have done throughout section~\ref{maxlines}.   In particular, this means that we can ignore the Witten effect and related subtleties in Maxwell theory with non-vanishing~$\theta$-angle. 

The choice of~$SU(2)$ representation $\mathbf{D}$ and the number $N_{f}$ of Weyl fermion flavors determines the possible flavor symmetries of the model~\eqref{nonabactiondef}, as well as whether the model is bosonic (with spacetime symmetry~$SO(4)$ or fermionic (with spacetime symmetry~$\text{Spin}(4)$). Together these comprise the connected 0-form symmetry group~$G^{(0)}$ of the theory. At long distances, we thus obtain Maxwell theory enriched by $G^{(0)}$ symmetry, exactly as discussed in section \ref{maxlines}.  By tracking the fractionalized~$G^{(0)}$ charges of the resulting Wilson and 't Hooft lines we will deduce the flavor and gravitational anomalies of the theory. 

Since anomalies are robust under all symmetry-preserving deformations, the anomalies we find can be understood to belong to a broader class of models than those defined precisely by \eqref{nonabactiondef}.  Among such anomaly-preserving deformations are changes of the scalar potential $V(\Phi)$.  We can therefore deform $V(\Phi)$ to gap out the scalar~$\Phi$ at a parametrically high scale and effectively remove it from the theory.  Thus the anomalies we describe below also apply to the QCD-like theory consisting of only the gauge fields and massless fermions.\footnote{~Here we do not mean that the theory is necessarily vector like.}  

A related comment is that we can tune the potential $V(\Phi)$ to make the Higgsing scale large relative to any strong-coupling scale of the non-Abelian gauge theory \eqref{nonabactiondef}.  In our analysis below, we are therefore free to assume that the model remains weakly coupled along the entire RG flow into Maxwell theory. 

\subsection{Yukawa Couplings, Symmetries, and Bundles}

Let us analyze the properties of the theories \eqref{nonabactiondef} in more detail.  The transformation properties of the fermions are explicitly written as
\begin{equation}\label{ferms}
\Psi_{\alpha}^{iA}~, \hspace{.3in}\alpha=1,2 ~(\text{left-handed Weyl spinor})~, \hspace{.2in}i\in \mathbf{D}~,\hspace{.2in}A=1,2, \cdots, N_{f}~.
\end{equation}
Similarly, we denote the scalar Higgs field by $\Phi_{a}$, where $a=1,2,3$ takes values in the adjoint representation of $SU(2)$.

To formulate the Yukawa couplings we must pair two fermions with the scalar.   This requires an invariant tensor $T$ which may be viewed as a singlet in the following tensor product of~$SU(2)$ representations,
\begin{equation}
T_{ij}^{a}\sim \mathbf{1} \subset \mathbf{D}\times \mathbf{D}\times \mathbf{3}~.
\end{equation}
The symmetry properties of the $i,j$ indices depend on whether $\mathbf{D}$ has even or odd dimension~$D$: if $D$ is even, then $T_{ij}^{a}$ is symmetric; and if $D$ is odd, then $T_{ij}^{a}$ is anti-symmetric. From this we can deduce the structure of the Yukawa couplings,
\begin{equation}
{\mathscr L}_\text{Yukawa}=\lambda_{AB} \, \varepsilon^{\alpha \beta} \,  T_{ij}^{a} \, \Phi_{a} \Psi_{\alpha}^{iA}\Psi_{\beta}^{jB}~, \hspace{.5in}\lambda_{AB} =\begin{cases} - \lambda_{BA} & \text{anti-symmetric if } D~\text{odd}~, \\
 \lambda_{BA} & \text{symmetric if } D~\text{even}~.
\end{cases}
\end{equation}

\subsubsection{0-Form Symmetries}

We will assume that the matrix~$\lambda_{AB}$ of Yukawa couplings is non-degenerate and therefore defines an invariant tensor of the flavor symmetry $G^{(0)}$.  The flavor symmetry is thus given by
\begin{equation}\label{simnot}
D~\text{odd} \; \longrightarrow \; G^{(0)}_\text{flavor} \sim  USp(N_{f} = 2n_f)~, \qquad D~\text{even} \; \longrightarrow \; G^{(0)}_\text{flavor} \sim SO(N_{f})~.
\end{equation}
Note that the non-degeneracy of $\lambda_{AB}$ implies that when $D$ is odd, the number of Weyl fermion flavors $N_{f}$ is even. We therefore write~$N_f = 2n_f$, where~$n_f$ is the number of Dirac fermion flavors. The Yukawa couplings can then be taken to be canonical invariant tensors~$\Omega_{AB} = -\Omega_{BA}$ and~$\delta_{AB} = \delta_{BA}$ for~$USp(2n_f)$ and~$SO(N_f)$ respectively, and up to a phase $\chi$,
\begin{equation}\label{chidef}
D~\text{odd} \; \longrightarrow \; \lambda_{AB}=\chi \Omega_{AB}~, \qquad D~\text{even} \; \longrightarrow \; \lambda_{AB}=\chi \delta_{AB}~, \qquad |\chi|=1~.
\end{equation}
In particular, a suitable choice of $\chi$ is sufficient to ensure that in the IR the effective $\theta$-angle vanishes, as assumed in~\eqref{irthetazero}.

The $\sim$ notation in \eqref{simnot} above indicates that we have not yet been precise about the possible discrete quotients that enter the 0-form symmetry~$G^{(0)}$.  Let us now describe these in greater detail:
\begin{itemize}
\item $D$~odd: There is an identification between the $\mathbb{Z}_{2}$ center of $USp(2n_f)$ and fermion number $(-1)^{F} \in \text{Spin}(4)$.  Therefore the total connected symmetry group, including flavor and spacetime symmetries, is given by
\begin{equation}\label{diagspinsp}
G^{(0)} = \frac{\text{Spin}(4)\times USp(2 n_{f})}{\mathbb{Z}_{2}}~.
\end{equation}
This allows us to determine the allowed bundles for~$G^{(0)}$ background fields, as in the discussion around \eqref{seq}.  Specifically, we may consider spacetimes~$M_4$ which are non-spin, and hence have non-zero $w_{2}(M_{4})\in H^{2}(M_{4},\mathbb{Z}_{2})$, as well as flavor bundles of $USp(2n_{f})/\mathbb{Z}_{2}$ which do not lift to $USp(2n_{f})$, and hence have non-zero $w_{2}(USp(2n_{f}))/\mathbb{Z}_{2})\in H^{2}(M_{4},\mathbb{Z}_{2})$.  However the diagonal quotient in \eqref{diagspinsp} implies the following constraint,
\begin{equation}
w_{2}(M_{4})=w_{2}(USp(2n_{f})/\mathbb{Z}_{2})~.
\end{equation}

\item $D$~even:  $(-1)^{F}$ is identified with the center of the $SU(2)$ gauge group and hence acts trivially on all gauge invariant local operators. Therefore the theory is bosonic, i.e.~the spacetime symmetry is~$SO(4) = \text{Spin}(4)/\IZ_2$. Further, if $N_{f}$ is even, the element $(-1)\in SO(N_{f})$ is also identified with the center of the gauge group.  Therefore the total connected symmetry group in the two cases is as follows,
\begin{equation}\label{sosymgroups}
N_{f} = 2 n_f~\text{even}:~G^{(0)} = SO(4) \times\frac{ SO(2 n_f)}{\mathbb{Z}_{2}}~, \qquad N_{f}~\text{odd}:~G^{(0)} = SO(4) \times SO(N_{f})~.
\end{equation}

Again we can translate these precise global symmetry groups into a discussion about the allowed background bundles. For reasons explained below, we will almost always assume that $N_{f}=2n_{f}$ is even, and hence we will focus on this case. Since the spacetime symmetry is~$SO(4)$, we may consider spacetimes that are non-spin, and hence have non-zero $w_{2}(M_{4})\in H^{2}(M_{4},\mathbb{Z}_{2})$.  We may also consider bundles of  $SO(2n_{f})/\mathbb{Z}_{2}$ that do not lift to bundles of the simply-connected universal cover $\text{Spin}(2n_{f})$. The nature of this obstruction depends on whether $n_{f}$ is even or odd, which controls the center $Z$ of $\text{Spin}(2n_{f})$,
\begin{equation}\label{centersplit}
Z(\text{Spin}(2n_{f}))=\begin{cases} \mathbb{Z}_{2, L}\times \mathbb{Z}_{2, R} & n_{f}~\text{even}~,\\
\mathbb{Z}_{4} & n_{f}~\text{odd}~.
\end{cases}
\end{equation}
\begin{itemize}
\item For $n_{f}$ even, there is a pair of $\mathbb{Z}_{2}$-valued obstruction classes characterizing bundles of $SO(2n_{f})/\mathbb{Z}_{2}$ that do not lift to bundles of $\text{Spin}(2n_{f})$.  We denote them by a pair:
\begin{equation}\label{wrldef}
w_{2,L}(SO(2n_{f})/\mathbb{Z}_{2})\in  H^{2}(M_{4},\mathbb{Z}_{2})~, \hspace{.3in}w_{2,R}(SO(2n_{f})/\mathbb{Z}_{2})\in  H^{2}(M_{4},\mathbb{Z}_{2})~.
\end{equation}
For future reference, we note that the obstruction to lifting to a bundle of $SO(2n_{f})$ is the sum $w_{2,L}+w_{2,R}$. Since there is no quotient in~\eqref{sosymgroups} that relates the flavor and spacetime symmetries, the characteristic classes in~\eqref{wrldef} are not related to~$w_2(M_4)$. 
  
\item For $n_{f}$ odd, there is a single $\mathbb{Z}_{4}$-valued obstruction class characterizing bundles of $SO(2n_{f})/\mathbb{Z}_{2}$ that do not lift to bundles of $\text{Spin}(2n_{f})$.  We denote it as
\begin{equation}
w_{2}(SO(2n_{f})/\mathbb{Z}_{2}) \in H^{2}(M_{4},\mathbb{Z}_{4})~.
\end{equation}
Now the obstruction to lifting to an $SO(2n_{f})$ bundle is the class $2w_{2}(SO(2n_{f})/\mathbb{Z}_{2}) \in H^{2}(M_{4},\mathbb{Z}_{2})$ (i.e.\ the reduction modulo $2$ of $w_{2}(SO(2n_{f})/\mathbb{Z}_{2})$).  As for~$n_f$ even, there is no relation between the characteristic classes for flavor and spacetime background bundles. 

\end{itemize}
\end{itemize}
\subsubsection{1-Form Symmetry}

In addition to the ordinary symmetries discussed above, there can also be an electric 1-form symmetry associated with the~$\IZ_2$ center of the~$SU(2)$ gauge group~\cite{Gaiotto:2014kfa}. When the~$SU(2)$ representation of the fermion is even dimension~$D$, the center  of the gauge group acts non-trivially on dynamical matter fields and hence such matter fields can screen all Wilson lines.  Correspondingly there is no 1-form symmetry.  By contrast, in the case where $D$ is odd, the center of the gauge group acts trivially on all dynamical matter fields and the fundamental Wilson line cannot be screened.  This leads to a $\mathbb{Z}_{2}^{(1)}$ 1-form symmetry.  In summary:
\begin{equation}\label{one-form sym}
D~\text{odd} \; \rightarrow \;  \mathbb{Z}_{2}^{(1)}~, \qquad D~\text{even} \; \rightarrow \; \text{none}~.
\end{equation}
For $D$ odd, the background field associated with the~$\IZ_2^{(1)}$ symmetry is a class $B_{\text{UV}}^{(2)} \in H^{2}(M_4,\mathbb{Z}_{2})$.

\subsection{The Quantum Numbers of Magnetic Monopoles}\label{qnmmsec}

We are now ready to understand the RG flow into Maxwell theory and to extract the fractional quantum numbers of the Wilson and 't Hooft lines in the IR.  As explained above, we assume that the Higgsing from~$SU(2)$ to~$U(1)$ happens at a high scale, so that the non-Abelian gauge theory is weakly coupled and amenable to semiclassical analysis.  Of course, the resulting anomalies are independent of this assumption.

In a weakly-coupled theory the Wilson line can be identified by examining the elementary electrically charged objects in the theory. These are either the dynamical charged fields in the theory, or the fundamental $SU(2)$ Wilson line in models with 1-form symmetry (see~\eqref{one-form sym}). Either way, determining the fractionalized~$G^{(0)}$ quantum numbers of the minimal~$W_1$ Wilson line is straightforward. Note also the following general constraint: the~$W_2 \sim W_1^2$ Wilson line represents the electrically charged~$W$-bosons of the Higgsed~$SU(2)$ theory, which do not carry any fractional~$G^{(0)}$ quantum numbers in the models we study. Therefore the fractional~$G^{(0)}$ quantum numbers of~$W_1$ are at most~$\half$.

The quantum numbers of IR 't Hooft lines are more subtle. The~$SU(2)$ gauge theory with non-zero vev for the Higgs field $\Phi^{a}$ has finite mass, topologically stable soliton particles of magnetic charge~$M = 1$.\footnote{~In the simplest case of a BPS monopole, these solitons satisfy the Bogomolny equations~$F \sim *d\Phi$~.} At long distances, they flow to the minimal 't Hooft line~$H_1$ of Maxwell theory. 

As was first understood in~\cite{Jackiw:1975fn}, the fractional quantum numbers of the monopole under~$G^{(0)}$ are determined by quantizing the fermion zero modes bound to the monopole. The number of such zero modes, as well as their transformation properties under the unbroken symmetries, can be deduced from suitable index theorems for the Dirac operator (see for instance \cite{Callias:1977kg,Moore:2014jfa}).  To state the result of the index theorem it is convenient to decompose the~$SU(2)$ gauge representation $\mathbf{D}$ (with dimension~$D = |{\bf D}|$) under the unbroken~$U(1)$ Cartan subgroup that governs the IR Maxwell theory. Hence this decomposition gives the electric charges of the resulting massive fermions,
\begin{equation}\label{decompab}
\mathbf{D}=(D-1)\oplus (D-3)\oplus \cdots \oplus -(D-3)\oplus -(D-1)~.
\end{equation}
For the minimal monopole of magnetic charge~$M=1$, the index theorem implies that each positive electric charge sector appearing above contributes a number of real fermion zero modes equal to the electric charge.  Therefore the total number of real fermion zero modes on the~$M = 1$ monopole is
\begin{equation}\label{indexthm}
\{\#~{\rm of} ~\IR\text{  zero-modes}\}= \begin{cases}\frac{N_{f}(D^{2}-1)}{4}& D~ \text{odd}~, \\
& \\ 
\frac{N_{f} D^{2}}{4}& D~ \text{even}~.
\end{cases}
\end{equation}
In all cases we study, gauge anomaly cancellation restricts~$N_f$ and~$D$ so that the number~$2 \mathcal{N}$ of real zero modes is always even. This is consistent with general expectations, as for instance recently discussed in~\cite{Sato:2022vii}. 

The index theorem also implies that the fermion zero modes associated to a sector of fixed electric charge in the decomposition \eqref{decompab} transform in an irreducible representation of the $SU(2)_\text{rot}$ little group of rotations  of the massive monopole.\footnote{~In fact, the $SU(2)_\text{rot}$ that leaves the monopole solution invariant is diagonally embedded into the~$SU(2)$ gauge group and the ordinary rotation group.}  As a representation of $\text{flavor}\times \text{spin}$, the fermion zero modes therefore transform as follows,
\begin{equation}\label{indexthm}
\mathbf{N_{f}} \otimes \left(\mathbf{D -1} \oplus \mathbf{D -3} \oplus \cdots \oplus \mathbf{2} \right)~~ D~\text{odd}~, \qquad \mathbf{N_{f}} \otimes \left(\mathbf{D -1} \oplus \mathbf{D -3} \oplus \cdots \oplus \mathbf{1} \right)~~D~\text{even}~.
\end{equation}
Note that these are real representations (i.e.\ they are representations defined over $\mathbb{R}$), which is required for compatibility with the reality of the fermion zero modes.  Explicitly, when~$D$ is odd then the~$SU(2)_\text{rot}$ representations are pseudoreal, while the $\mathbf{N_{f} = 2 n_f}$ is the pseudoreal fundamental representation of the~$USp(2 n_{f})$ flavor symmetry.  Meanwhile, in the case of even $D$ both the $SU(2)_\text{rot}$ representations and the fundamental $\mathbf{N_{f}}$ of $SO(N_{f})$ are real.

We can now deduce the spin and flavor representation of the monopole (and hence the fractional~$G^{(0)}$ charges of the 't Hooft line) by quantizing the zero modes above to obtain the possible monopole ground states. If there are~$2 \mathcal{N}$ real fermion zero modes, their anti-commutation relations give rise to a Clifford algebra and the resulting fermionic Fock space is a Dirac spinor representation of the associated $\text{Spin}(2\mathcal{N})$ group.  Our task will be to understand how this Dirac spinor decomposes into projective representations of~$G^{(0)}$. In general, we will find that the monopole states do transform projectively under~$G^{(0)}$, and hence the 't Hooft line in the IR Maxwell theory carries fractional~$G^{(0)}$ charges. 

An important subtlety concerns the electric charge of the fermion zero modes.  In general, the moduli space of classical solutions that contains the basic~$M = 1$ monopole solution includes an~$S^1$ factor that is acted on by large $U(1)$ gauge transformations.  Quantizing this classical moduli space of solutions gives rise to bare dyons, whose electric and magnetic charges in our conventions are given by $(E,M)=(2n,1)$ with $n\in\mathbb{Z}$.  These dyons can be viewed as bound states of~$(2,0)$ W-bosons with the basic~$(0,1)$ monopole.

To this tower of bare dyons we must incorporate the fermion zero modes to obtain the true spectrum of particles with unit magnetic charge. It is common in such a discussion to combine the fermion zero modes with suitable bare dyons to make real fermions.  This has been implicit in our discussion above.  

However, since the bare dyons only carry even electric charge, the real fermion zero modes discussed above retain a well-defined electric charge modulo 2.  As with much of our discussion, the physical consequences of this observation depend on the parity of $D$:
\begin{itemize}
\item $D$ odd: all electric charges of fermions in the Maxwell phase are even.  Hence by combining with suitable bare dyon excitations we can assume that all fermion zero modes have vanishing electric charge.
\item $D$ even: all electric charges of fermions in the Maxwell phase are odd. Hence the action of a fermion zero mode changes the parity of electric charge.  Therefore quantizing the zero modes leads to the Fock space associated to an elementary~$(0,1)$ monopole and that associated to an elementary~$(1,1)$ dyon.
\end{itemize}
A closely related discussion concerns the chiral spinors that make up the Dirac spinor of $\text{Spin}(2\mathcal{N})$.  The Dirac spinor always decomposes into two chiral spinors distinguished by their eigenvalues $\pm1$ under a chirality operator that anti-commutes with all zero modes. The physical meaning of this chirality decomposition therefore also depends on the parity of $D$:
\begin{itemize}
\item $D$ odd: The chirality operator acts on the fermion zero modes as a global symmetry, $(-1)^{F}$, and hence the monopole Hilbert space always consists of both bosons and fermions. 
\item $D$ even:  The chirality operator acts on the fermion zero modes as electric charge modulo~2.  Therefore, the decomposition of the Fock space into eigenstates of the chirality operator specifies the electric charge~$E$ modulo~$2$.   
\end{itemize}
Both of these statements are familiar from the study of monopoles in~$SU(2)$ gauge theories with ${\mathcal N} = 2$ supersymmetry, see for instance  \cite{Seiberg:1994rs,Seiberg:1994aj,Harvey:1996ur}, where gauginos furnish adjoint fermions with $D=3$, while fundamental hypermultiplets give matter fermions with~$D = 2$. As explained in these references, the fact that the monopole and the dyon must be quantized as chiral and anti-chiral spinors of~$\text{Spin}(2 {\cal N})$ when~$D$ is even is necessary for the consistency of dyon-antimonople scattering.

\subsection{$SU(2)$ with $N_f  =1$ and Even $D$}

Our first example is $SU(2)$ gauge theory with a single Weyl fermion $\Psi_{\alpha}^{i}$ in an even-dimensional representation $\mathbf{D}$ of the gauge group.  This example was discussed in \cite{Wang:2018qoy} and our analysis closely parallels the discussion there.  It is well-known \cite{Witten:1982fp} that this theory has a gauge anomaly that renders it inconsistent if~$D \equiv 2~(\text{mod }4)$. Comparing with the number ${D^2 \over 4}$ of real fermion zero modes in \eqref{indexthm}, we also see that this number would be odd. An odd number of fermion zero modes cannot be consistently quantized to describe a single monopole and should be viewed as an inconsistency, as was recently emphasized in~\cite{Sato:2022vii}.  In this subsection, we therefore assume that $D \equiv 0~(\text{mod } 4)$.

 We follow the general procedure outlined above and Higgs the model to $U(1)$ using a real adjoint scalar $\Phi^{a}$.  Since this model is bosonic and has no internal symmetries, our task is to investigate the possible charge fractionalization of the~$G^{(0)} = SO(4)$ spacetime symmetry, by examining whether the Wilson or 't Hooft lines transform as fermions under~$(-1)^F \in {\widetilde G}^{(0)} = \text{Spin}(4)$.  

Since $\Psi_{\alpha}^{i}$ is a fermion, we see that the basic Wilson line~$W_1$ should also be a fermion. We thus conclude that
\begin{equation}\label{spincfrac}
Q_{e}=  \frac{\sigma_{e}}{2}=\frac{1}{2} \quad \Longrightarrow \quad b_{e}^{(2)}=w_{2}(M_{4})~.
\end{equation}
This can also be seen from the fact that the dynamical~$SU(2)$ gauge field of the UV theory is really an~$SO(3)$ gauge field that is constrained to satisfy~$w_2(SO(3)) =w_2(M_4)$. This is because the UV gauge and global symmetries are~$(SU(2) \times \text{Spin}(4))/\IZ_2$, with a common~$\IZ_2$ center. Higgsing to the Cartan leads to a version of Maxwell theory based on a dynamical $\text{Spin}_c$ connection. As was explained around~\eqref{spincab}, this is equivalent to activating the background fields~\eqref{spincfrac} in ordinary Maxwell theory based on a~$U(1)$ connection. 

Next we turn to the quantum numbers of the basic~$M = 1$ monopole. As explained around~\eqref{indexthm}, the index theorem guarantees the existence of real fermion zero modes transforming as follows under the~$SU(2)_\text{rot}$ little group of the monopole, 
\begin{equation}\label{spinors1}
{\bf 1}\oplus{\bf 3}\oplus{\bf 5}\oplus...\oplus \bf{ D-1}~.
\end{equation}
Let us label each of the summands above by their dimension $d$, so that~$d = 1, 3, \ldots, D-1$.  In each summand we have a collection $\chi_{d}^{m}$ of zero modes with
\begin{equation}
\{\chi_{d}^{m}, \chi_{d}^{n}\}=2\delta^{mn}~, \qquad (\chi_{d}^{m})^{\dagger}=\chi_{d}^{m}~, \qquad m=1,2, \ldots, d~.
\end{equation}
By contrast, the zero modes from different summands anti-commute. Note that the total number $\frac{|D|^{2}}{4}$ of real zero modes is in fact even, because we are assuming that~$D \equiv 0~(\text{mod }4)$. Therefore the fermionic Fock space can be constructed in a standard fashion.

It is simplest to decompose the representation \eqref{spinors1} under the Cartan subgroup $U(1)_\text{rot}\subset SU(2)_\text{rot}$ and to form complex combinations of the fermion zero modes that carry definite $U(1)_\text{rot}$ charge. We use non-negative integer (rather than half-integer) weights~$J \in \IZ_{\geq 0}$ to label representations of  $U(1)_\text{rot}$. We will denote a complex zero mode carrying $U(1)_\text{rot}$ charge~$J$ by~$a^{\dagger}_{J}$, while its Hermitian conjugate $a_{J}$ carries charge~$-J$. Since all~$SU(2)_\text{rot}$ representations in \eqref{spinors1} have integer spin, $J$ will only take on even integer values. Taking into account the degeneracies arising from the representation \eqref{spinors1}, we thus arrive at the following fermion zero-mode algebra:
\begin{equation}
\{a_{J}^{i}, a_{J}^{\dagger j}\}=\delta^{ij}~, \qquad J=0, 2, \ldots, D-2~, \qquad i,j =\begin{cases}1, 2, \ldots , \frac{1}{2}(D-J) & J>0~, \\
1, 2, \ldots, \frac{1}{4}D& J=0~.
\end{cases}
\end{equation}  

We now use these modes to construct a Fock space: we introduce a Clifford vacuum $|\Omega \rangle$ that is annihilated by all~$a_{-J}^i$ lowering operators. In general, the Clifford vacuum $|\Omega \rangle$ itself carries some $U(1)_\text{rot}$ charge $J_{\Omega} \in \IZ_{\geq 0}$. The Fock space is then spanned by states obtained by acting on $|\Omega \rangle$ with all combinations of raising operators $a^{\dagger i}_{J}$,
\begin{equation}
|\Omega \rangle~, \hspace{.1in}a^{\dagger i_{1}}_{J_{1}}|\Omega \rangle~, \hspace{.1in}a^{\dagger i_{2}}_{J_{2}}a^{\dagger i_{1}}_{J_{1}}|\Omega \rangle~, ~~\ldots\hspace{.1in}
\end{equation}
Note that since the modes all have even~$U(1)_\text{rot}$ charge, the parity of $J_{\Omega}$ controls whether the Fock space has even charge and is thus a boson, or whether it has odd charge and is therefore a fermion. 

To answer this question, we must fix the~$U(1)_\text{rot}$ charge~$J_\Omega$ of the Clifford vacuum. There is a standard way to do this: we examine the state of largest $U(1)_\text{rot}$ charge in the Fock space, which is obtained by acting with all raising operators on the Clifford vacuum. Hence this state has~$U(1)_\text{rot}$ charge
\begin{equation}
J_\text{max}=J_{\Omega}+\sum_{J~\text{even } > \, 0}^{D-2}\frac{J(D-J)}{2}=J_{\Omega}+\frac{D(D^{2}-4)}{24}~.
\end{equation}
In order for the Fock space to form complete $SU(2)_\text{rot}$ representations it is necessary that the largest $U(1)_\text{rot}$ charge be equal and opposite the smallest $U(1)_\text{rot}$ charge.  Since the latter comes from the Clifford vacuum itself, this means that~$J_\text{max} = - J_\Omega$, which requires~$J_\Omega$ to be negative,  
\begin{equation}\label{jomega}
J_\Omega = - \frac{D(D^{2}-4)}{48} < 0~.
\end{equation}
Recalling that $D \equiv 0~(\text{mod } 4)$, we conclude from~\eqref{jomega} that the parity of $J_{\Omega}$ only depends on $D$ modulo $8$.  Thus: 
\begin{equation}
D \equiv 0~(\text{mod } 8) \; \Longrightarrow \; |J_{\Omega}|~ \text{even} \; \leftrightarrow \; \text{Bosonic}~, \qquad D \equiv 4~(\text{mod } 8) \; \Longrightarrow \; |J_{\Omega}|~ \text{odd} \; \leftrightarrow \; \text{Fermionic}~.
\end{equation}

Note that, inline with the discussion at the end of section \ref{qnmmsec}, the Fock space constructed above is naturally split: there is a sector of even electric charge describing the~$(E, M) = (0,1)$ magnetic monopole, and a sector of odd electric charge describing the elementary~$(1,1)$ dyon.  These sectors correspond to starting from the Clifford vacuum $|\Omega \rangle$ and acting with an even numbers of fermion zero modes for the monopole, or with an odd number of zero modes for the dyon.  From the discussion above, we conclude that the monopole and the dyon have the same statistics. Using the spin indicators introduced around \eqref{spinfusion} we have therefore deduced that
\begin{equation}\label{dmod48spins}
\sigma_{m}=\sigma_{d}=\begin{cases} 
0 &D \equiv 0~(\text{mod } 8)~,\\
1 & D \equiv 4~(\text{mod } 8)~.\\
\end{cases}
\end{equation}
Note that this is consistent with the fusion constraint \eqref{spinfusion}.  

In particular, we see from~\eqref{dmod48spins} that the case~$D \equiv 4~(\text{mod } 8)$ produces all-fermion electrodynamics, as reviewed in section \ref{sec:allfed}.  Therefore this theory has a discrete gravitational anomaly. This can be understood by examining the magnetic 2-form background, 
\begin{equation}
b_{m}^{(2)}=\frac{D}{4}w_{2}(M_4)\in H^{2}(M_{4},\mathbb{Z}_{2})~, \qquad D \equiv 0~(\text{mod } 4)~.
\end{equation}
Since the electric 2-form background is~$b_e^{(2)} = w_2(M_4)$ (see~\eqref{spincfrac}), we conclude that there is an 't Hooft anomaly of the form 
\begin{equation}
\mathcal{A}=\frac{i \pi D}{4} \int_{M_5}  w_{2}(M_5)\cup \beta_2(w_{2}(M_5))=\frac{i \pi D}{4}  \int_{M_5} w_{2}(M_5)\cup w_{3}(M_5)~.
\end{equation} 
This reproduces~\eqref{gravanom} when $D \equiv 4~(\text{mod } 8)$. Although we have deduced this anomaly using the renormalization group flow to Maxwell theory, it survives any symmetry-preserving deformation.  In particular, this anomaly is present in the theory of gauge fields and fermions where the scalar is removed. In this way we make contact with the results of \cite{Wang:2018qoy}. 

\subsection{$SU(2)$ with Fundamental Fermions}

Our next class of examples is $SU(2)$ gauge theory with $N_{f}$ Weyl fermions transforming in the fundamental (doublet) representation of the~$SU(2$) gauge group.  To avoid a gauge anomaly of Witten type \cite{Witten:1982fp}, the number of doublets must be even, $N_{f}=2n_{f}$.\footnote{~This can also be seen by requiring the number of real fermion zero modes around a~$M = 1$ monopole, which by~\eqref{indexthm} is~$N_f {D^2 \over 4} = N_f$, to be even (see \cite{Sato:2022vii}).}   As discussed in \eqref{sosymgroups}, the Yukawas reduce the connected 0-form symmetry of the theory to
\begin{equation}
G^{(0)} = SO(4) \times\frac{ SO(2n_{f})}{\mathbb{Z}_{2}}~.
\end{equation}
According to \eqref{centersplit}, the possible fractionalized charges~$\vec Q$ of line defects (defined in \eqref{fracdef}) depend on the parity of $n_{f}\,$:  
\begin{itemize}
\item $n_{f}$ even.  A fractionalization $\vec Q$ is a triple:
\begin{equation}
\vec Q=\left(\frac{q_{L}}{2}, \frac{q_{R}}{2}, \frac{\sigma}{2}\right)~,
\end{equation}
where $q_{L},$ $q_{R}$, and $\sigma$ are all valued in $\mathbb{Z}_{2}$. Here~$(q_L, q_R)$ specify the charge of the line under the $\mathbb{Z}_{2, L}\times \mathbb{Z}_{2, R}$ center of $\text{Spin}(2n_{f})$, while~$\sigma$ specifies whether it is a boson or fermion. 
\item $n_{f}$ odd.  A fractionalization $\vec Q$ is a pair:
\begin{equation}
\vec Q=\left(\frac{q}{4},  \frac{\sigma}{2}\right)~,
\end{equation}
where now $q\in \mathbb{Z}_{4}$ specifies the charge of the line under the center of $\text{Spin}(2n_{f})$, while~$\sigma \in \IZ_2$ specifies the statistics of the line. 
\end{itemize}

\subsubsection{Quantum Numbers of Wilson Lines} 

In the Maxwell gauge theory obtained after Higgsing, the quantum numbers of the fundamental~$W_1$ Wilson line are dictated by those of the dynamical~$SU(2)$ doublet fermion in the UV non-Abelian gauge theory. This line is therefore a fermion in the vector representation of $SO(2n_{f})$, and hence is fractionalized with respect to both the flavor and spacetime symmetries.  This fixes the fractionalized charges~$\vec Q_e$ of the Wilson line, 
\begin{equation}\label{qefund}
\vec Q_{e}=\begin{cases}
\left(\frac{1}{2}, \frac{1}{2}, \frac{1}{2}\right) & n_{f}~\text{even}~, \\
\left({2 \over 4} = \frac{1}{2}, \frac{1}{2}\right) & n_{f}~\text{odd}~, \\ 
\end{cases}
\end{equation}
Here we have used the fact that when $n_{f}$ is even, the vector representation of~$SO(2n_f)$ has equal charges~$q_L= q_R = 1$ under the two factors in the~$\IZ_{2, L} \times \IZ_{2, R}$ center of~$\text{Spin}(2n_f)$. By contrast, when~$n_f$ is odd, the vector representation is characterized by having charge~$q = 2$ under the $\mathbb{Z}_{4}$ center of~$\text{Spin}(2n_f)$.

We now apply~\eqref{backgroundid2} to deduce  the electric 2-form background gauge field~$b_e^{(2)}$ in the two cases:
\begin{itemize}
\item $n_f$ even. In this case all fractionalizations are~$\IZ_2$-valued, so that~$b_e^{(2)} \in H^2(M_4, \IZ_2)$. Using the fractional charges~$q_L = q_R = \sigma = 1$ in~\eqref{qefund}, we find
\begin{equation}\label{befundeven}
 b_{e}^{(2)} = w_{2,L}\left(\frac{SO(2n_{f})}{\mathbb{Z}_{2}}\right)+w_{2,R}\left(\frac{SO(2n_{f})}{\mathbb{Z}_{2}}\right)+w_{2}(M_4) \in H^2(M_4, \IZ_2) \qquad (n_f \text{ even})~.
\end{equation}

\item $n_f$ odd. Now the flavor symmetry is fractionalized in~$\IZ_4$, while spin is fractionalized in~$\IZ_2$, (i.e.~it follows from the discussion around~\eqref{backgroundid2} that~$N=4$) and hence~$b_e^{(2)} \in H^2(M_4, \IZ_4)$. The fractionalized charges~$q = 2$ and~$\sigma = 1$ from~\eqref{qefund} then imply that
\begin{equation}
b_e^{(2)} = 2w_{2}\left(\frac{SO(2n_{f})}{\mathbb{Z}_{2}}\right)+2 w_{2}(M_4) \in H^2(M_4, \IZ_4) \qquad (n_f \text{ odd})~.
\end{equation}
Here we are interpreting~$2 w_2(M_4)$ as~$\IZ_4$ class, as explained around~\eqref{longexii}. For future reference, it is useful to rewrite this as follows,
\begin{equation}\label{befundodd}
b_e^{(2)}  = 2 \left( \left[w_{2}\left(\frac{SO(2n_{f})}{\mathbb{Z}_{2}}\right)\right]_2 + w_2(M_4)\right)~,
\end{equation}
where we use the notation~$[\cdots]_2$ to denote the reduction of a~$\IZ_4$ class modulo 2. This shows that~$b_e^{(2)}$ is in fact a~$\IZ_2$ class (namely the class that appears inside the parentheses in~\eqref{befundodd}), which is embedded into~$H^2(M_4, \IZ_4)$ via multiplication by 2.

\end{itemize}

\noindent The fact that~$b_e^{(2)}$ always turns out to be a~$\IZ_2$ class, even though it is a priori~$\IZ_4$ when~$n_f$ is odd, is non-trivially consistent with the general constraint mentioned at the beginning of section~\eqref{qnmmsec}: the fractionalization class of the~$E = 1$ Wilson line can be at most~$\IZ_2$, since two such lines fuse into an~$E = 2$ line representing the~W-bosons, which carry no fractional~$G^{(0)}$ charges.  

\subsubsection{Quantum Numbers of 't Hooft Lines} 

We now turn to the quantum numbers of the basic~$H_1$ 't Hooft line. The index theorem discussed around equation \eqref{indexthm} implies that each Weyl fermion gives a single real fermion zero mode~$\chi^{A}$ of vanishing~$SU(2)_\text{rot}$ spin, which leads to the following Clifford algebra for the zero modes,
\begin{equation}
\{\chi^{A}, \chi^{B}\}=2\delta^{AB}~, \qquad (\chi^A)^\dagger = \chi^A~, \qquad A,B=1,2, \ldots, 2n_{f}~.
\end{equation}
The resulting Fock space is therefore completely bosonic and transforms as a Dirac spinor of the $\text{Spin}(2n_{f})$ flavor symmetry.

To correctly identify the quantum numbers of the 't Hooft line, we must remember the discussion at the end of section \ref{qnmmsec}: the fermion zero modes carry odd electric charge, and hence the chiral and anti-chiral spinors of $\text{Spin}(2n_{f})$ that make up the Dirac spinor constructed above have electric charge~$E = 0$ (corresponding to the~$(0,1)$ monopole), and~$E = 1$ (corresponding to the~$(1,1)$ dyon), respectively. Therefore their transformations under $\text{Spin}(2n_{f})$ are distinct, though both are spacetime bosons. 

We can now summarize the fractional charges~$\vec Q_m$ and~$\vec Q_d$ of the~$(0,1)$ monopole and the~$(1,1)$ dyon as follows: when~$n_f$ is even, a (anti-) chiral spinor only transforms under the (right) left~$\IZ_2$ factor in the $\mathbb{Z}_{2, L}\times \mathbb{Z}_{2, R}$ center of $\text{Spin}(2n_{f})$. This implies that
\begin{equation}\label{oddnfm}
n_{f}~\text{even}: ~~\vec Q_{m}=\left(\frac{1}{2},0, 0\right)~, \;  \vec Q_{d}=\left(0,\frac{1}{2}, 0\right) \; \Longrightarrow \; b_{m}^{(2)}=w_{2,L}\left(\frac{SO(2n_{f})}{\mathbb{Z}_{2}}\right) \in H^2(M_4, \IZ_2)~.
\end{equation}
By contrast, for odd $n_{f}$ the two chiral spinors transform with charge $\pm1$ (mod 4) under the~$\IZ_4$ center of $\text{Spin}(2n_{f})$, so that
\begin{equation}\label{evennfm}
n_{f}~\text{odd}: ~~\vec Q_{m}=\left(\frac{1}{4},0\right)~, \; \vec Q_{d}=\left(\frac{3}{4}, 0\right) \; \Longrightarrow \; b_{m}^{(2)}=w_{2}\left(\frac{SO(2n_{f})}{\mathbb{Z}_{2}}\right) \in H^2(M_4, \IZ_4)~.
\end{equation}
A simple consistency check of these results is that they are consistent with fusion of line defects, which implies
\begin{equation}
\vec Q_{e}+\vec Q_{m}=\vec Q_{d}+\left(\vec{0},\frac{1}{2}\right)~,
\end{equation}
i.e.\ the flavor quantum numbers add, while the statistics is shifted to account for the angular momentum in the electromagnetic field.

\subsubsection{'t Hooft Anomaly} 

We can now assemble our results to determine the 't Hooft anomaly for the~$G^{(0)}$ in these theories, by substituting into the general formula \eqref{anomform}, or equivalently~\eqref{discreteanom}.  As always, the answer depends on the parity of $n_{f}$:
\begin{itemize}
\item $n_{f}$ even: substituting the~$\IZ_2$ classes~$b_{e, m}^{(2)}$ in~\eqref{befundeven} and~\eqref{oddnfm} into~\eqref{discreteanom} (with~$N = 2$), we find the following anomaly, 
\begin{equation}\label{afirstformnfodd}
\mathcal{A}=i\pi \int_{M_{5}}\left(w_{2,L}\left(\frac{SO(2n_{f})}{\mathbb{Z}_{2}}\right)+w_{2,R}\left(\frac{SO(2n_{f})}{\mathbb{Z}_{2}}\right)+w_{2}(M_5)\right)\cup \beta_2 \left(w_{2,L}\left(\frac{SO(2n_{f})}{\mathbb{Z}_{2}}\right)\right)~.
\end{equation}
This expression can be usefully simplified by noting that on an orientable five-manifold we have the following relation for any $X\in H^{2}(M_{5},\mathbb{Z}_{2})$ (see for instance appendix D of~\cite{Kapustin:2017jrc}),\footnote{~The relation \eqref{xsqxrel} can be deduced by noting that, with $\mathbb{Z}_{2}$ coefficients, the map $X\mapsto X\cup \beta_2(X)$ is linear and evaluates to a top form.  Therefore, by Poincar\'{e} duality, it is expressible by a cup product with a universal characteristic class.}
\begin{equation}\label{xsqxrel}
 \int_{M_{5}}X\cup \beta_2 (X)= \int_{M_{5}}X\cup w_{3}(M_{5})= \int_{M_{5}}w_{2}(M_{5})\cup \beta_2 (X)~.
\end{equation}
Applying this relation to \eqref{afirstformnfodd} simplifies expression for the anomaly, 
\begin{eqnarray}\label{afinalnfeven}
n_f \text{ even}: \quad \mathcal{A} &=&i\pi \int_{M_{5}}w_{2,R}\left(\frac{SO(2n_{f})}{\mathbb{Z}_{2}}\right)\cup \beta_2 \left(w_{2,L}\left(\frac{SO(2n_{f})}{\mathbb{Z}_{2}}\right)\right)\\
& =&i\pi \int_{M_{5}}w_{2,L}\left(\frac{SO(2n_{f})}{\mathbb{Z}_{2}}\right)\cup \beta_2 \left(w_{2,R}\left(\frac{SO(2n_{f})}{\mathbb{Z}_{2}}\right)\right)~. \nonumber
\end{eqnarray}
Note in particular that this anomaly is symmetric if we exchange the two chiral spinors of~$\text{Spin}(2 n_f)$. 

\item $n_{f}$ odd: now we substitute the~$\IZ_4$ classes~$b_{e, m}^{(2)}$ in~\eqref{befundodd} and~\eqref{evennfm} into~\eqref{discreteanom} (with~$N = 4$), to find the following anomaly, 
\begin{equation}
\mathcal{A}={i\pi \over 2} \int_{M_{5}}\beta_4 \left(2 \left( \left[w_{2}\left(\frac{SO(2n_{f})}{\mathbb{Z}_{2}}\right)\right]_2 + w_2(M_5)\right)\right) \cup w_{2}\left(\frac{SO(2n_{f})}{\mathbb{Z}_{2}}\right)~.
\end{equation}
Although this anomaly is superficially~$\IZ_4$ valued, it is in fact actually valued in~$\IZ_2$. This can be seen by integrating~$\beta_4$ by parts, so that
\begin{equation}\label{afinalnfodd}
n_f \text{ odd}: \quad \CA = i \pi \int_{M_5}  \left( \left[w_{2}\left(\frac{SO(2n_{f})}{\mathbb{Z}_{2}}\right)\right]_2 + w_2(M_5)\right) \cup \left[ w_3\left(\frac{SO(2n_{f})}{\mathbb{Z}_{2}}\right) \right]_2~.
\end{equation}
Here~$w_3\left(SO(2n_{f})/ \mathbb{Z}_{2}\right) = \beta_4 \left(w_{2}\left( SO(2n_{f}) / \mathbb{Z}_{2}\right)\right) \in H^3(M_5, \IZ_4)$. We pause to emphasize that $[w_3(SO(2n_f)/\IZ_2)]_2 \neq \beta_2 \left( [w_2(SO(2n_f)/\IZ_2)]_2\right)$. If this were the case, it would follow from~\eqref{xsqxrel} that the anomaly vanishes, but as we argue below it does not in fact vanish. 
\end{itemize}

We can subject the anomaly formulas in~\eqref{afinalnfeven} and~\eqref{afinalnfodd} to the following consistency checks:
\begin{itemize}
\item {\it Mass Deformations:} If we restrict the~$SO(2n_{f})/\mathbb{Z}_{2}$ flavor symmetry to a subgroup of the form $SO(2n_{f}-4)/\mathbb{Z}_{2}$, the center of the universal cover is unmodified, and hence so are the obstruction classes.  This means that the anomalies described above remain visible in $SO(2n_{f}-4)/\mathbb{Z}_{2}$. We can therefore add masses to remove Weyl fermions in groups of four without modifying the anomalies.  This explains why the latter only depend on~$N_f$ modulo 4, or equivalently on~$n_f$ modulo 2.\footnote{~The case $n_{f}=1$ is an exception.  In reducing from $SO(4k+2)/\mathbb{Z}_{2}$ to $SO(2)/\mathbb{Z}_{2}$ the center of the universal cover is modified from~$\IZ_4$ to~$\mathbb R$, and as a result the discrete anomalies discussed above are trivialized.} 

\item {\it Restriction to~$SO(2n_f)$:} Even though the~$\IZ_2$ quotient in the~$SO(2n_f)/\IZ_2$ flavor symmetry gives us access to a larger set of background bundles, we are free to only consider bundles that lift to~$SO(2n_f)$. Now there is a single class~$w_2(SO(2n_f)) \in H^2(M_4, \IZ_2)$, which obstructs further lifts to~$\text{Spin}(2n_f)$. Since the Wilson line is a vector of~$SO(2n_f)$, its flavor quantum numbers are not fractionalized, but it is still a fermion, so that~$b_e^{(2)} = w_2(M_4) \in H^2(M_4, \IZ_2)$. The monopole is a boson and a spinor of~$SO(2n_f)$, so that~$b_m^{(2)} = w_2(SO(2n_f))$. Therefore the 't Hooft anomaly restricted to~$SO(2n_f)$ bundles takes the form
\begin{equation}\label{soanom}
\CA[SO(2n_f)] = i \pi \int_{M_5} w_2(M_5) \cup w_3(SO(2n_f)) = i \pi \int_{M_5} w_2(SO(2n_f)) \cup w_3(SO(2n_f))~.
\end{equation}
In the last step we have used~\eqref{xsqxrel}, which holds for~$\IZ_2$ classes. Note that this formula is uniform in~$n_f$. 

It is straightforward to check that \eqref{soanom} is reproduced by the more refined~$SO(2n_f)/\IZ_2$ anomalies upon restriction to~$SO(2n_f)$ bundles. When~$n_f$ is even, this restriction means setting~$w_{2, L}(SO(2n_f)/\IZ_2) = w_{2, R}(SO(2n_f)/\IZ_2) = w_2(SO(2n_f))$, so that~\eqref{afinalnfeven} correctly reduces to the second expression on the right-hand side of~\eqref{soanom}. When~$n_f$ is odd, we have the relation~$w_2(SO(2n_f)/\IZ_2) = 2 w_2(SO(2n_f)) \in H^2(M_4, \IZ_4)$. This has vanishing reduction modulo~2, so that the anomaly~\eqref{afinalnfodd} reduces to
\begin{equation}
n_f \text{ odd}: \qquad \CA[SO(2n_f)] = i \pi \int_{M_5} w_2(M_5) \cup \left[\beta_4\left(2 w_2(SO(2n_f))\right)\right]_2~.
\end{equation}
However, it can be shown that~$\left[\beta_4\left(2 X\right)\right]_2 = \beta_2 \left(X \right) \in H^3(M_5, \IZ_2)$ for any~$X \in H^2(M_5, \IZ_2)$, so that the anomaly correctly reduces to the first expression in~\eqref{soanom}. 
\end{itemize}

\subsubsection{Comparison with the Perturbative Chiral Anomaly of~$SU(2)$ QCD}

The anomalies derived in \eqref{afinalnfeven} and \eqref{afinalnfodd} (and also~\eqref{soanom}) are robust under any symmetry-preserving deformations.  One such deformation involves turning off the Yukawa couplings and gapping out the adjoint scalar field $\Phi^{a}$ via its potential~$V(\Phi)$.  This leads to QCD with~$SU(2)$ gauge group and $N_f = 2n_{f}$ massless fundamental Weyl fermions. (Equivalently, $n_f$ massless fundamental Dirac fermions.) Here we explain how to understand the anomalies \eqref{afinalnfeven} and \eqref{afinalnfodd} from the point of this~$SU(2)$ QCD theory. 

We first describe the global symmetries and 't Hooft anomalies of~$SU(2)$ QCD.  Its flavor symmetry is larger than the~$SO(2n_{f})/\mathbb{Z}_{2}$ preserved by the Yukawa couplings to $\Phi$.  Specifically, we have an enhancement $SO(2n_{f})/\mathbb{Z}_{2}\rightarrow SU(2n_{f})/\mathbb{Z}_{2}$ where the $\mathbb{Z}_{2}$ quotient of $SU(2n_{f})$ acts via multiplication by $(-1)$.  The analog of the symmetry group \eqref{sosymgroups} is then given by
\begin{equation}\label{susymgroups}
SO(4) \times \frac{SU(2n_{f})}{\mathbb{Z}_{2}}~.
\end{equation}
Hence the allowed bundles have two independent discrete characteristic classes,  
\begin{equation}\label{discreet}
w_{2}(M_{4})\in H^{2}(M_{4}, \mathbb{Z}_{2})~,\hspace{.3in}w_{2}\left(\frac{SU(2n_{f})}{\mathbb{Z}_{2}}\right)\in H^{2}(M_{4}, \mathbb{Z}_{2})~.
\end{equation}

The global symmetry~\eqref{susymgroups} admits various possible 't Hooft anomalies:
\begin{itemize}
\item {\it Perturbative chiral anomaly:}  Since the symmetry $SU(2n_{f})/\mathbb{Z}_{2}$ acts chirally on the fermions, it has a standard perturbative (i.e.~triangle) anomaly.  The anomaly action $\mathcal{A}$ is a five-dimensional Chern-Simons term that can be defined by extending the five-manifold $M_{5}$ to the boundary of a six-manifold $M_{6}$ (with~$\partial M_6 = M_5$) and using descent, so that 
\begin{equation}\label{chernclass}
\mathcal{A}\supset 2\pi i \int_{M_{6}}c_{3}\left(\frac{SU(2n_{f})}{\mathbb{Z}_{2}}\right)~, \qquad \partial M_6 = M_5~.
\end{equation}
Since this anomaly is fully characterized by a three-point function of~$SU(2n_f)$ flavor currents, it can be detected without paying attention to the $\mathbb{Z}_{2}$ quotient in $SU(2n_{f})/\mathbb{Z}_{2}$.

\item {\it Discrete anomalies:}  Using the discrete characteristic classes~\eqref{discreet}, we can formulate two possible  anomalies characterized by anomaly coefficients $x,y \in \mathbb{Z}_{2}$,
\begin{equation}\label{adiscretesu2}
\mathcal{A} \supset \pi i x\int_{M_{5}}w_{2}(M_{5})\cup w_{3}(M_{5})+\pi i y\int_{M_{5}}w_{3}(M_{5})\cup w_{2}\left(\frac{SU(2n_{f})}{\mathbb{Z}_{2}}\right)~.
\end{equation}
Here we have made use of the relation \eqref{xsqxrel} to reduce the number of possibly anomaly terms. 
\end{itemize}
\noindent One can think of the full anomaly of~$SU(2)$ QCD as a sum of the standard perturbative anomaly~\eqref{chernclass}, and the discrete anomaly \eqref{adiscretesu2}.  A priori we do not know the~$\IZ_2$ coefficients~$x, y$ that appear in~\eqref{adiscretesu2}, but we will now argue that they must both vanish, $x = y = 0$. 

To see this, we consider mass deformations of~$SU(2)$ QCD. In the notation of~\eqref{ferms}, such a deformation is characterized by an anti-symmetric mass matrix $m_{AB}$,
\begin{equation}
{\mathscr L}_m = m_{AB} \Psi_{\alpha}^{iA}\Psi_{\beta}^{jB}\varepsilon^{\alpha\beta}\varepsilon_{ij} + (\text{h.c.})~, \qquad  m_{AB}=m_{[AB]}~.
\end{equation}
Choosing a non-degenerate mass matrix~$m_{AB} \sim \Omega_{AB}$ thus breaks the flavor symmetry as follows,
\begin{equation}
SU(2n_{f})/\mathbb{Z}_{2} \; \longrightarrow \; USp(2n_{f})/\mathbb{Z}_{2}~,
\end{equation}
with~$\Omega_{AB}$ the invariant symplectic tensor of~$USp(2n_{f})$. Notice that the~$\mathbb{Z}_{2}$ quotient of the $SU(2n_{f})$ and~$USp(2n_f)$ groups is by their common center. Therefore the discrete characteristic classes of the two groups are related by a simple restriction, 
\begin{equation}\label{bundrest}
w_{2}\left(\frac{SU(2n_{f})}{\mathbb{Z}_{2}}   \right)\bigg|_{USp(2n_{f})/\mathbb{Z}_{2}}=w_{2}\left(\frac{USp(2n_{f})}{\mathbb{Z}_{2}}\right)~.
\end{equation} 
Here the left-hand side denotes value of the~$SU(2n_f)/\IZ_2$ bundle discrete characteristic class when the structure group of the bundle is reduced to~$USp(2n_{f})/\mathbb{Z}_{2}$, while the right-hand side is the  intrinsic discrete characteristic class of an~$USp(2n_{f})/\mathbb{Z}_{2}$ bundle. From now on we will no longer distinguish them. 

A consequence of \eqref{bundrest} is that, unlike the perturbative anomaly \eqref{chernclass} (which vanishes after a mass deformation), the possible discrete anomalies \eqref{adiscretesu2} are preserved under mass deformations. If such discrete anomalies are present in~$SU(2)$ QCD, they must therefore also be present in the pure~$SU(2)$ Yang-Mills theory without matter that we can flow to by making the masses large. Moreover, we can dial the phases of the masses to engineer pure~$SU(2)$ Yang-Mills theory with~$\theta = 0$. This theory is anomaly free, and is believed to flow to a trivially gapped confining phase in the IR. We conclude that the discrete anomaly coefficients $x,y$ in~\eqref{adiscretesu2} must vanish in~$SU(2)$ QCD. Therefore the only anomaly of interest is the familiar chiral 't Hooft anomaly \eqref{chernclass}.

We now argue this anomaly is responsible for the discrete fractionalization anomalies~\eqref{afinalnfeven} and \eqref{afinalnfodd} that we found by deforming the theory to pure Maxwell theory via an adjoint Higgs field and Yukawa couplings. We must therefore understand how \eqref{chernclass} behaves under restriction to a subgroup $SO(2n_{f})/\mathbb{Z}_{2} \subset SU(2n_f)/\IZ_2$, distinguishing between even and odd~$n_f\,$:

\begin{itemize}

\item $n_{f}$ even.  Thanks to the general discussion below \eqref{afinalnfodd} about reducing~$n_f$ in units of two via anomaly-preserving mass terms, it is sufficient to consider the minimal case $n_{f}=2$. Then the symmetries in question enjoy exceptional isomorphisms,
\begin{equation}
\text{no Yukawas:}~~SU(4)/\mathbb{Z}_{2}\cong SO(6)~, \hspace{.3in}\text{with Yukawas:}~~SO(4)/\mathbb{Z}_{2}\cong SO(3)_L \times SO(3)_R~.
\end{equation}
Under the identification $SU(4)/\mathbb{Z}_{2}\cong SO(6)$ the third Chern class defining the perturbative anomaly \eqref{chernclass} is identified with the Euler class,
\begin{equation}
c_{3}\left(\frac{SU(4)}{\mathbb{Z}_{2}}\right)=e_{6}\left(SO(6)\right)\in H^{6}(M_{6},\mathbb{Z})~.
\end{equation}
When reduced to structure group $SO(3)_L\times SO(3)_R$, the Euler class obeys a product formula, 
\begin{equation}
e_{6}\left(SO(6)\right)|_{SO(3)_L\times SO(3)_R}=e_{3}(SO(3)_L)\cup e_{3}(SO(3)_R)~,
\end{equation}
where above $e_{3}(SO(3))\in H^{3}(M_{6},\mathbb{Z})$ denotes the Euler class of each factor and the subscripts $L,R$ label the factors.  However, the Euler class $e_{3}(SO(3))$ of an $SO(3)$ bundle is equal to the third integral Stiefel-Whitney class of the bundle, 
\begin{equation}
e_{3}(SO(3))=W_{3}(SO(3))~.
\end{equation}
Combining the steps above thus leads to the following reduction formula,
\begin{equation}
c_{3}\left(\frac{SU(2n_{f})}{\mathbb{Z}_{2}}\right)\bigg|_{SO(3)_L\times SO(3)_R}=W_{3}(SO(3)_L)\cup W_{3}(SO(3)_R)~.
\end{equation}
Notice that the right-hand side above is an integral cohomology class, but that it is torsion.  Indeed the third integral Stiefel-Whitney class of an $SO(3)$ bundle is equal to the integral Bockstein~$\beta$ of the second Stiefel-Whitney class,\footnote{~In other words, $\beta$ is the connecting homomorphism resulting from the coefficient sequence $0\rightarrow \mathbb{Z}\rightarrow \mathbb{Z} \rightarrow \mathbb{Z}_{2}\rightarrow 0$.}
\begin{equation}
\beta(w_{2}(SO(3)))=W_{3}(SO(3)) \in H^{3}(M_{6}, \mathbb{Z}) \; \Longrightarrow \; 2W_{3}(SO(3))=0\in H^{3}(M_{6}, \mathbb{Z})~.
\end{equation}
The physical consequence of the anomaly reducing to a torsion class is that it can no longer be detected by studying local current correlation functions, since such local anomalies can never be torsion.

Although the local anomaly vanishes after reducing the structure group to~$SO(2n_f)/\IZ_2$, there is still a torsion class remaining.  We can evaluate it using the inflow definition \eqref{chernclass}, noting that reducing modulo 2 gives $\left[W_{3}(SO(3))\right]_2 =w_{3}(SO(3))$, and recalling that the Bockstein map is a derivation.  Therefore we may integrate it by parts to obtain\footnote{~The last step is justified by using the fact that the Bockstein of a class $X$ with $\mathbb{Z}_{2}$ coefficients can be obtained by picking a lift $\tilde{X}$ to an integral chain and evaluating $\frac{1}{2}\delta \tilde{X}=\beta(X).$}
\begin{eqnarray}
\mathcal{A} & = & 2\pi i \int_{M_{6}}c_{3}\left(\frac{SU(4)}{\mathbb{Z}_{2}}\right)\bigg|_{SO(3)_L\times SO(3)_R}\nonumber\\
&= & 2\pi i \int_{M_{6}}W_{3}(SO(3)_L)\cup W_{3}(SO(3)_R)\label{match}\\
& = &i \pi \int_{M_{5}}w_{2}(SO(3)_L)\cup \beta_2 \left(w_{2}(SO(3)_R)\right)\nonumber~.
\end{eqnarray}
Finally, we identify 
\begin{equation}
w_{2}(SO(3)_L)=w_{2,L}\left(\frac{SO(4)}{\mathbb{Z}_{2}}\right)~, \qquad w_{2}(SO(3)_R)=w_{2,R}\left(\frac{SO(4)}{\mathbb{Z}_{2}}\right)~,
\end{equation}
and conclude that \eqref{match} exactly matches the anomaly \eqref{afinalnfeven} derived from the fractionalized quantum numbers of lines in the Maxwell phase.

\item $n_{f}$ odd. In this case, we wish to show that the chiral anomaly reduces to~\eqref{afinalnfodd} upon restricting~$SU(2n_f)/\IZ_2 \; \rightarrow SO(2n_f)/\IZ_2$. For simplicity, we only carry out this analysis For~$SU(2n_f) \rightarrow SO(2n_f)$ bundles, in which case we must show that the chiral anomaly reduces to~\eqref{soanom}. It is known \cite{Toda} that
\begin{equation}
c_3(SU(2n_f))\big|_{SO(2n_f)} = W_3(SO(2n_f)) \cup W_3(SO(2n_f))~, 
\end{equation}
where~$W_3(SO(2n_f))$ is the integral third Stiefel-Whitney class of the~$SO(2n_f)$ bundle. Since this class is 2-torsion, $W_3(SO(2n_f)) \cup W_3(SO(2n_f))$ need not vanish in integer cohomology. Technically, we can now follow essentially the same steps as in~\eqref{match} to reduce the chiral anomaly~\eqref{chernclass} to~$SO(2n_f)$. This yields the same answer as in~\eqref{match}, except that~$w_{2}(SO(3)_L)$ and~$w_{2}(SO(3)_R)$ are both replaced by~$w_2(SO(2n_f))$. The resulting answer precisely matches the fractionalization anomaly in~\eqref{soanom}.

\end{itemize}

\subsection{$SU(2)$ with Adjoint Fermions}\label{su2adjoints}

As a final example, we consider~$SU(2)$ gauge theory with~$N_f = 2n_{f}$ Weyl fermions in the adjoint representation of the gauge group (i.e.~we are considering~$D= 3$) and a real adjoint scalar $\Phi$.  As in our previous discussion we will tune the scalar potential so that $\Phi$ condenses.  The resulting IR theory is weakly coupled, and it contains Maxwell theory, so that we  can investigate the quantum numbers of its lines and the anomalies implied by them. 

Unlike our previous examples, the IR theory now also contains some weakly coupled fermions that are uncharged under the Maxwell gauge field.  These arise because the Yukawa couplings in~\eqref{diagspinsp}, for the~$D = 3$ adjoint case, do not give a mass to those adjoint fermions that are color-aligned with the vev of the adjoint Higgs field~$\Phi_a$. The IR theory thus contains~$2n_f$ gauge-neutral Weyl fermions that transform in the fundamental representation of the~$USp(2n_f)$ flavor subgroup of the full connected symmetry preserved by non-degenerate Yukawa couplings (see~\eqref{diagspinsp}),\footnote{~For~$n_f = 1,2$, this is familiar from~$\CN=2, 4$ supersymmetric gauge theories (see below). There, the neutral fermions are the massless gaugino superpartners of the IR Maxwell field on the Coulomb branch.} 
\begin{equation}
\frac{\text{Spin}(4)\times USp(2n_{f})}{\mathbb{Z}_{2}}~.
\end{equation}
Note that unlike the examples above, this theory is fermionic, i.e.~it has gauge-invariant local operators that are fermions. However, fermion number~$(-1)^F \in \text{Spin}(4)$ is identified with the central element~$-1 \in USp(2n_f)$. 

The presence of these gauge-neutral fermions is in fact required by anomaly matching: when $D = 3$ is odd, the~$SU(2)$ gauge theory in the UV has an 't Hooft anomaly of Witten type~\cite{Witten:1982fp} for the~$USp(2n_f)$ flavor symmetry, which cannot be matched by pure Maxwell theory.\footnote{~Roughly speaking, this is because the anomaly of Maxwell theory is always factorized, as in~\eqref{anomu1}, while the Witten anomaly is not. See~\cite{Garcia-Etxebarria:2017crf} for a related discussion.} In the IR, this anomaly is matched by the gauge-neutral Weyl fermions in the fundamental of~$USp(2n_f)$. Note that these fermions will not interfere with our analysis of the anomalies matched by the IR Maxwell theory below.

Additionally this model has an exact $\IZ_2^{(1)}$ electric 1-form symmetry (i.e.~a center symmetry).  This symmetry acts non-trivially on all Wilson lines in even-dimensional representations of~$SU(2)$ and prevents them from breaking.  We denote the associated background field by $B^{(2)}_\text{UV} \in H^{2}(M_{4},\mathbb{Z}_{2})$.  In the UV, this background field controls the 't Hooft flux of the~$SU(2)$ gauge theory. In other words, one is instructed to carry out the gauge theory path integral as if the gauge group were~$SO(3)$, subject to the constraint
\begin{equation}
w_2(SO(3)) = B^{(2)}_\text{UV} ~.
\end{equation}
Upon Higgsing to~$U(1)$, we find that the IR Maxwell gauge field~$\t a^{(1)}$ satisfies the following twisted flux-quantization condition,
\begin{equation}
\int_{\Sigma_2} {d \t a^{(1)} \over 2 \pi} \in \half \int_{\Sigma_2} B^{(2)}_\text{UV}  + \IZ~.
\end{equation}
As explained in section \ref{spincsec}, this theory has an alternative but equivalent presentation as a conventional~$U(1)$ connection~$a^{(1)}$, which satisfies \begin{equation}
\int_{\Sigma_2} {d  a^{(1)} \over 2 \pi} \in \IZ~,
\end{equation}
coupled to a flat, $2\pi$-periodic electric 2-form background gauge field~$B_e^{(2)}$ of the form
\begin{equation}\label{UVIRrel}
B_e^{(2)} = \pi B^{(2)}_\text{UV} ~. 
\end{equation}

Now let us describe the possible fractionalizations $\vec Q$ of  line defects.  To account for the 1-form symmetry and the relation \eqref{UVIRrel}, we generalize the discussion around \eqref{fracdef} to a pair that controls the charge under $\mathbb{Z}_{2}^{(1)}$ as well as the possible action of the $\mathbb{Z}_{2}$ in the common center of $\text{Spin}(4)\times USp(2n_{f})$.  Thus we write:
\begin{equation}
\vec Q=\left(\frac{q}{2},\frac{\sigma}{2}\right)~, \hspace{.2in} q, \sigma \in \mathbb{Z}_{2}~.
\end{equation}
Here~$q$ is given in terms of the electric charge $E$, 
\begin{equation}
q= E ~ \mod ~2~.
\end{equation}
If we denote by~$(-1)^{F_A}$ the central element~$(-1) \in USp(2n_f)$, then we can also write 
\begin{equation}\label{sigadj}
\sigma= \text{charge of}~(-1)^{F}+\text{charge of }~(-1)^{F_A}~.
\end{equation}
The fundamental Wilson line~$W_1$ of Maxwell theory arises from the UV Wilson line in the doublet of $SU(2)$ (not from the dynamical fermionic matter fields, all of which have even electric charge).  Therefore we have the following electric charge fractionalization and background field,
\begin{equation}\label{fraceadj}
\vec Q_{e}=\left(\frac{1}{2},0\right)\quad \Longrightarrow \quad b_{e}^{(2)}=B^{(2)}_\text{UV}~.
\end{equation}

Next we turn to the fundamental 't Hooft line~$H_1$, whose quantum numbers are fixed by those of the charge~$M = 1$ magnetic monopole.  The index theorem \eqref{indexthm} predicts~$2 \times 2n_f$ real zero modes
\begin{equation}
\gamma_\alpha^A \in \mathbf{2} \otimes  \left(\mathbf{2n_{f}}\right) ~.
\end{equation}
Here~$\alpha = 1,2$ is a spinor index under the spatial rotation group $SU(2)_\text{rot}$ under which the bosonic monopole solution is invariant, while~$A = 1, \ldots, 2n_f$ is a fundamental~$USp(2n_f)$ index. Note that the zero modes are invariant under the common centers of these symmetry groups, i.e.~they are not themselves fractionalized. This in turn implies that it is possible to consistently assign a unique fractionalization class to all states in the monopole Hilbert space. 

To construct the monopole  Hilbert space, we must quantize the zero modes, which satisfy
\begin{equation}
\left(\gamma^A_\alpha\right)^\dagger = \gamma_A^\alpha~, \qquad \{\gamma^A_\alpha, \gamma_B^\beta\} = 2 \delta^A_B \delta_\alpha^\beta~.
\end{equation}
The~$\gamma^A_\alpha$ can be viewed as Hermitian Dirac matrices of~$\text{Spin}(4n_f)$, and quantizing them yields the~$2^{2n_f}$-dimensional Dirac spinor representation of that group. To determine the fractionalization pattern, we must therefore decompose this Dirac spinor representation under the subgroup 
\begin{equation}\label{adjointdecomp}
SU(2)_\text{rot} \times USp(2n_f) \subset \text{Spin}(4n_f)~. 
\end{equation}
We first consider small values of~$n_f$, before proceeding to the general case:

\begin{itemize}

\item $n_{f}=1$.  For a single Dirac fermion the groups in \eqref{adjointdecomp} simplify to:
\begin{equation}
SU(2)_\text{rot} \times SU(2)_{A} \subset \text{Spin}(4)~.
\end{equation}
Then, $\gamma_\alpha^A$ is a $\text{Spin}(4)$ gamma matrix, with~$\alpha = 1,2$ and~$A = 1,2$ left-chiral and right-chiral~$SU(2)$ spinor indices.  Quantizing the zero modes leads to a 4-component Dirac spinor, which decomposes into chiral components:
\begin{equation}
(\mathbf{2}, \mathbf{1}) \oplus (\mathbf{1},\mathbf{2}) \in \text{Rep}\left(SU(2)_\text{rot} \times SU(2)_A\right)~.
\end{equation}
The~$(\mathbf{2}, \mathbf{1}) $ representation is a spin-$\half$ fermion that is neutral under the flavor symmetry. By contrast, the~$(\mathbf{1},\mathbf{2}) $ representation is a spin-0 boson transforming as a doublet under the~$SU(2)_{A}$ flavor symmetry.

Not coincidentally, this is precisely the representation content of the monopole hypermultiplet in pure~$SU(2)$ Yang-Mills theory with~${\cal N} = 2$ supersymmetry. This is because our model can be embedded into the supersymmetric theory, with~$SU(2)_A$ playing the role of the~$SU(2)_R$ symmetry that acts on the supercharges. Comparing the monopole quantum numbers to \eqref{sigadj}, we find that all states in the monopole multiplet have 
\begin{equation}
\sigma= \text{charge of}~(-1)^{F}+\text{charge of }~(-1)^{F_{A}}=1~.
\end{equation}
Therefore, the monopole fractionalization is:
\begin{equation}
\vec Q_{m}=\left(0,\frac{1}{2}\right)\quad \Longrightarrow \quad b_{m}^{(2)}=w_{2}(M_{4})=w_{2}\left(\frac{SU(2)_{A}}{\mathbb{Z}_{2}}\right)~.
\end{equation}

\item $n_f = 2$.  For a pair of Dirac fermions the groups in \eqref{adjointdecomp} simplify to: 
\begin{equation}
SU(2)_\text{rot} \times \text{Spin}(5)_{A} \subset \text{Spin}(8)~,
\end{equation}
where we have used the isomorphism $USp(4) \cong \text{Spin}(5)$.  Then~$\gamma_\alpha^A$ is a~$\text{Spin}(8)$ gamma matrix transforming in the~$\mathbf{8_v}$ vector representation. Quantizing the zero modes leads to a 16-component Dirac spinor, which decomposes into left- and right-chiral spinors~$\mathbf{8_s}$ and~$\mathbf{8_c}$ of~$\text{Spin}(8)$.

To decompose this representation under $SU(2)_\alpha \times \text{Spin}(5)_A$ it is convenient to use~$\text{Spin}(8)$ triality to exchange~$\mathbf{8_v} \leftrightarrow \mathbf{8_s}$, so that the gamma matrices transform as a left-chiral spinor of~$\text{Spin}(8)$.  Using the fact that $SU(2)_\text{rot}\cong \text{Spin}(3)_\text{rot}$, the chiral spinor decomposes into simultaneous spinors under  $SU(2)_\text{rot} \times \text{Spin}(5)_{A}$:
\begin{equation}
\mathbf{8_s}\sim\gamma_\alpha^A \rightarrow (\mathbf{2}, \mathbf{4}) \in \text{Rep}\left(SU(2)_\text{rot} \times \text{Spin}(5)_A\right)~.
\end{equation}
The same decomposition also applies to the~$\mathbf{8_c}$ anti-chiral spinor.  By contrast, the~$\mathbf{8_v}$ vector decomposes into a direct sum of vectors,
\begin{equation}
\mathbf{8_v} \rightarrow (\mathbf{3}, \mathbf{1}) \oplus (\mathbf{1},\mathbf{5}) \in \text{Rep}\left(\text{SU}(2)_\text{rot} \times \text{Spin}(5)_A\right)~.
\end{equation}
Thus the 16-dimensional Dirac spinor decomposes into the following representations:
\begin{equation}
\text{Dirac} = \mathbf{8_c} \oplus \mathbf{8_v} \rightarrow (\mathbf{2}, \mathbf{4}) \oplus (\mathbf{3}, \mathbf{1}) \oplus (\mathbf{1}, \mathbf{5}) \in \text{Rep}\left(SU(2)_{rot} \times \text{Spin}(5)_A\right)~.
\end{equation}
Note that this is exactly the representation content of a monopole vector multiplet on the Coulomb branch of~$SU(2)$ pure gauge theory with~${\cal N} = 4$ supersymmetry, into which our model can be embedded. In this case~$USp(4)/\IZ_2$ is identified with the~$SO(5)_R \subset SO(6)_R$ symmetry of the supersymmetric theory on its Coulomb branch. 

Comparing the monopole quantum numbers to \eqref{sigadj}, we see that all monopole states have
\begin{equation}
\sigma= \text{charge of}~(-1)^{F}+\text{charge of }~(-1)^{F_{A}}=0~.
\end{equation}
Therefore, the monopole fractionalization is:
\begin{equation}
\vec Q_{m}=\left(0,0\right)\quad \Longrightarrow \quad b_{m}^{(2)}=0~.
\end{equation}

\item General $n_{f}$.  We now show that for an arbitrary number~$n_f$ of adjoint Dirac fermions, the monopole fractionalization is
\begin{equation}\label{bmadjf}
\vec Q_{m}=\left(0,\frac{n_{f}}{2}\right) \quad \Longrightarrow \quad b_{m}^{(2)}=n_{f}w_{2}(M_{4})=n_{f}w_{2}\left(\frac{USp(2n_{f})}{\mathbb{Z}_{2}}\right)~.
\end{equation}
For general $n_{f}$, we can identify the bosonic and fermionic monopole state using the chirality of the spinors.  Indeed, the chirality operator $\gamma^*$ anticommutes with all mode operators and represents $(-1)^{F}$ acting on the Fock space.  We therefore have the identification of eigenspaces:
\begin{equation}
\gamma^{*}=+1 \leftrightarrow \text{bosonic monopoles}~, \hspace{.3in}\gamma^{*}=-1 \leftrightarrow \text{fermionic monopoles}~.
\end{equation}
Next we observe that for~$n_f$ odd the chiral spinors of~$\text{Spin}(4n_f)$ are pseudoreal, while for~$n_f$ even they are real. The bosonic states are in a real representation of~$SU(2)_\text{rot}$. Hence, pseudoreality (reality) of these states is dictated by whether they are charged (uncharged) under the center $(-1)^{F_A}$ of $USp(2n_{f})$.  Therefore we conclude that the bosonic states have~$(-1)^{F_A} = (-1)^{n_f}$. Similarly, the fermionic states are in a pseudoreal representation of~$SU(2)_\text{rot}$ and hence they have~$(-1)^{F_A} = (-1)^{n_f + 1}$.  We conclude that all monopole states satisfy
\begin{equation}
\sigma= \text{charge of}~(-1)^{F}+\text{charge of }~(-1)^{F_A}~=n_{f}~ \mod ~2~,
\end{equation}
completing the argument.

\end{itemize}

We can assemble our fractionalization data and the resulting background fields \eqref{fraceadj} and \eqref{bmadjf} to evaluate the 't Hooft anomaly of this theory. Using the general formula \eqref{anomform} we obtain a mixed 1-form/0-form anomaly characterized by the following inflow action,
\begin{equation}\label{adjresult}
\mathcal{A} = i \pi n_f \int_{M_5} B^{(2)}_\text{UV} \cup w_3(M_5) = i \pi n_f  \int_{M_5} B^{(2)}_\text{UV} \cup w_{3}\left(\frac{USp(2n_{f})}{\mathbb{Z}_{2}}\right)~. 
\end{equation}
As already mentioned above, the special case~$n_f =1$ also describes pure~$SU(2)$ gauge theory with~${\cal N} = 2$ supersymmetry, with the~$SU(2)_R$ symmetry of the supersymmetric theory playing the role of the~$SU(2)_A$ flavor symmetry above.   In that context the consequences of the anomaly were understood in \cite{Witten:1994cg,Witten:1995gf,Moore:1997pc}, and their interpretation as an anomaly was discussed in \cite{Cordova:2018acb}. If we add fundamental~${\cal N} =2$ hypermultiplets to the pure~$SU(2)$ gauge theory, the 1-form symmetry associated with~$B_\text{UV}^{(2)}$ is broken. However, we can still activate non-trivial~$(SU(2)_R \times \text{Spin}(4) ) / \IZ_2$ bundles, as long as we also set (see for instance~\cite{Aspman:2022sfj} for a recent discussion in the context of topological twisting) 
\begin{equation}
B_\text{UV}^{(2)} = w_2(M_4) = w_2\left({SU(2)_R \over \IZ_2}\right)~.
\end{equation}
Therefore the anomaly~\eqref{adjresult} (with~$n_f = 1$) survives the addition of fundamental hypermultiplets, and it takes the same form as the gravitational anomaly in all-fermion electrodynamics discussed in section~\ref{sec:allfed}. This is particularly intuitive if we add a single fundamental hypermultiplet, since the resulting theory realizes the original Argyres-Douglas SCFT on its Coulomb branch~\cite{Argyres:1995jj,Argyres:1995xn}. The Argyres-Douglas point enjoys a~$\IZ_3$ symmetry that acts via the order-three electric-magnetic duality transformation~$ST \in SL(2, \IZ)$. It therefore treats the fundamental electric, magnetic, and dyonic particles (all of which become massless at the SCFT point) completely symmetrically. As discussed in section~\ref{sec:allfed},  this is only possible in all-fermion electrodynamics. 

As with any anomaly, the result \eqref{adjresult} is robust under continuous symmetry preserving deformations.  Therefore we must find the same result in the $SU(2)$ adjoint QCD theory, which does not have the scalar field $\Phi$, and where the adjoint fermions are massless.

It is straightforward to interpret the anomaly \eqref{adjresult} in this context. To do so we must identify the background fields. The one-form symmetry background $B^{(2)}_\text{UV}$ is simply identified with the 1-form center symmetry background in $SU(2)$ adjoint QCD.  Meanwhile, the 0-form symmetry is enhanced in the absence of Yukawa couplings,
\begin{equation}
\frac{USp(2n_{f})}{\mathbb{Z}_{2}}\quad \rightarrow \quad\frac{SU(2n_{f})}{\mathbb{Z}_{2}}~.
\end{equation}
And under restriction the discrete characteristic classes are identified as:
\begin{equation}\label{bundrest2}
w_{2} \left(\frac{SU(2n_{f})}{\mathbb{Z}_{2}}\right)\bigg|_{\frac{USp(2n_{f})}{\mathbb{Z}_{2}}}=w_{2}\left(\frac{USp(2n_{f})}{\mathbb{Z}_{2}}\right)~,
\end{equation} 
Comparing with \eqref{adjresult}, we immediately conclude that $SU(2)$ adjoint QCD has the same anomaly, now re-interpreted as anomaly for the larger flavor symmetry, 
\begin{equation}\label{adjresult2}
\mathcal{A} = i \pi n_f \int_{M_5} B^{(2)}_\text{UV} \cup w_3(M_5) = i \pi n_f  \int_{M_5} B^{(2)}_\text{UV} \cup w_{3}\left(\frac{SU(2n_{f})}{\mathbb{Z}_{2}}\right)~. 
\end{equation}

\section*{Acknowledgements}

We thank J.~Harvey, K.~Intriligator, M.~Martone, and E.~Nardoni for discussions. DB, CC, and TD, acknowledge support from the Simons Collaboration on Global Categorical Symmetries. In addition CC is supported by the US Department of Energy DE-SC0009924. TD is supported by a DOE Early Career Award under DE-SC0020421 and the Mani L. Bhaumik Presidential Chair in Theoretical Physics at UCLA.

\bibliography{biblio}
\bibliographystyle{utphys}

\end{document}